\documentclass{emulateapj}

\usepackage{psfig}
\usepackage{amsmath}
\usepackage{amssymb}
\usepackage{graphicx}
\usepackage{appendix}

\newcommand{\lSect}[1]{{\label{sec:#1}}}
\newcommand{\lFig}[1]{{\label{fig:#1}}}
\newcommand{\lEq}[1]{{\label{eq:#1}}}
\newcommand{\lTab}[1]{{\label{tab:#1}}}
\def\gtaprx {\lower .1ex\hbox{\rlap{\raise .6ex\hbox{\hskip .3ex
	{\ifmmode{\scriptscriptstyle >}\else
		{$\scriptscriptstyle >$}\fi}}}
	\kern -.4ex{\ifmmode{\scriptscriptstyle \sim}\else
		{$\scriptscriptstyle\sim$}\fi}}}
\def\ltaprx {\lower .1ex\hbox{\rlap{\raise .6ex\hbox{\hskip .3ex
	{\ifmmode{\scriptscriptstyle <}\else
		{$\scriptscriptstyle <$}\fi}}}
	\kern -.4ex{\ifmmode{\scriptscriptstyle \sim}\else
		{$\scriptscriptstyle\sim$}\fi}}}
\newcommand{\FIGFF}[2]{{\ref{fig:#2}{#1}}}

\newcommand{\FIG}[2]{{Fig.~\FIGFF{#1}{#2}}}
\newcommand{\Fig}[1]{{\FIG{}{#1}}}

\newcommand{\Sectff}[1]{{\ref{sec:#1}}}
\newcommand{\Sect}[1]{{\S~\Sectff{#1}}}

\newcommand{\Eqref}[1]{{\ref{eq:#1}}}

\newcommand{\eqff}[1]{{\Eqref{#1}}}

\newcommand{\eq}[1]{{eq.~\eqff{#1}}}

\newcommand{\Tab}[1]{{Table \ref{tab:#1}}}
\newcommand{\Msun}{\mbox{$M_\odot$}}

\def\gtaprx {\lower .1ex\hbox{\rlap{\raise .6ex\hbox{\hskip .3ex
	{\ifmmode{\scriptscriptstyle >}\else
		{$\scriptscriptstyle >$}\fi}}}
	\kern -.4ex{\ifmmode{\scriptscriptstyle \sim}\else
		{$\scriptscriptstyle\sim$}\fi}}}
\def\ltaprx {\lower .1ex\hbox{\rlap{\raise .6ex\hbox{\hskip .3ex
	{\ifmmode{\scriptscriptstyle <}\else
		{$\scriptscriptstyle <$}\fi}}}
	\kern -.4ex{\ifmmode{\scriptscriptstyle \sim}\else
		{$\scriptscriptstyle\sim$}\fi}}}

\begin{document}

\title{Models for the Unusual Supernova \MakeLowercase{i}PTF14\MakeLowercase{hls}}

\author{S. E. Woosley\altaffilmark{1}}

\altaffiltext{1}{Department of Astronomy and Astrophysics, University
  of California, Santa Cruz, CA 95064; woosley@ucolick.org}

\begin{abstract} 
Supernova iPTF14hls maintained a bright, variable luminosity for more
than 600 days, while lines of hydrogen and iron in its spectrum had
different speeds, but showed little evolution. Here several varieties
of models are explored for iPTF14hls-like events. They are based upon
circumstellar medium (CSM) interaction in an ordinary supernova,
pulsational pair-instability supernovae (PPISN), and magnetar
formation. Each is able to explain the enduring emission and
brightness of iPTF14hls, but has shortcomings when confronted with
other observed characteristics.  The PPISN model can, in some cases,
produce a presupernova transient like the one observed at the site of
iPTF14hls in 1954. It also offers a clear path to providing the
necessary half solar mass of material at $\sim 5 \times 10^{16}$ cm
for CSM interaction to work, and can give an irregular light curve
without invoking additional assumptions. It explains the 4000 km
s$^{-1}$ seen in the iron lines, but without additional energy input,
strains to provide enough matter to explain the nearly constant 8000
km s$^{-1}$ velocity seen in H$_{\alpha}$. Magnetar models can also
explain many of the observed features, but give a smooth light curve
and may require an evolving magnetic field strength. Their dynamics
may be difficult to reconcile with the observation of slow-moving
hydrogen at late times.  The various models predict different spectral
characteristics and a remnant that, today, could be a black hole,
magnetar, or even a star. Further observations and calculations of
radiation transport will narrow the range of possibilities.
\end{abstract}


\section{INTRODUCTION}
\lSect{intro}

\MakeLowercase{i}PTF14\MakeLowercase{hls} was discovered by the iPTF
survey in September, 2014 \citep{Arc17}. Though initially sparsely
sampled, the failure of the supernova to decline marked it for more
intensive investigation. In January, 2015 it was determined to be a
Type II supernova. For the next year, the supernova displayed an
irregular light curve with multiple episodes of brightening during
which the luminosity varied by about 50\%. The total energy emitted in
light during the first 600 days was about $2.2 \times 10^{50}$ erg,
making iPTF14hls a luminous, but not particularly ``superluminous''
supernova. There is also some evidence of a supernova-like transient
having happened at the same site in 1954.  The long duration of
iPTF14hls precludes the supernova from being a purely recombination
event like most Type IIp supernovae. The required mass and energy are
too large. Nor is a radioactive energy source the explanation. No
radioactivity with the appropriate half life is produced with
sufficient abundance in any model.

One is thus left with two possibilities: 1) iPTF14hls was
collisionally powered, its light coming from shells of matter that
collided over a long period of time, possibly augmented by a central
source; or 2) the event was an extreme example, in terms of duration
and spectrum, of magnetar-illuminated supernovae. Both possibilities
were suggested by \citet{Arc17}, with the underlying cause for the
shell ejections in case 1 attributed to a pulsational-pair instability
supernova \citep[e.g.][]{Woo07a,Woo17}.

Here, both possibilities are explored. First, the outcome of an
ordinary supernova interacting with a dense wind is considered
(\Sect{csm}). Given a free choice of wind parameters, one can easily
find an approximate fit to the global light curve and the high
velocity (hydrogenic) component of the spectrum, though the origin of
the lower velocity iron lines is less clear in a one-dimensional model
\citep[though see][and \Sect{hybrid}]{And17}. Even more problematic,
the reason why a star would lose roughly a solar mass of high
velocity material during the last few decades of its life is not
obvious unless the star dies as a PPISN.

\Sect{ppisn} thus considers PPISN explanations for iPTF14hls. The
models are of three sorts: 1) those that might be capable of producing
both the 1954 and 2014 transients, but which leave a stellar core that
has not yet collapsed (\Sect{1954}); 2) slightly lighter stars where
the pulsing activity is restricted to the last decade of the star's
life, and which experience iron-core collapse while the light curve is
still in progress (\Sect{alone}); and 3) hybrid models that invoke an
asymmetric explosion with one component due to the collapse of the
iron core to a compact object (\Sect{hybrid}). Each has strengths and
weaknesses. All are able to explain the approximate duration and
brightness of the light curve and, since each can occur in nature, may
ultimately appear as iPTF14hls-like supernovae, if not as iPTF14hls
itself.  Each PPISN model has difficulty, though, explaining the high
velocity (8000 km s$^{-1}$) H$_{\alpha}$ component of iPTF14hls.

Magnetar models are considered in \Sect{magnetar}. It is not difficult
to find a two parameter fit that approximately tracks the overall
bolometric light curve. The very bright initial display of the
magnetar is masked by the overlying star and adiabatically degraded
alleviating a concern voiced by \citet{Arc17}. For the 20 \Msun
\ supernova model considered, most of the pulsar energy is invisible
for the first 100 days. Models with greater explosion energies
and more mass loss give higher speeds that may be necessary to explain
the spectrum. If the same neutron star is responsible for accelerating
the helium core to 4000 km s$^{-1}$ and making the light curve, magnetic
field decay must be invoked. While magnetar models are potentially
successful at explaining many characteristics of iPTF14hls, they would
not give the spectroscopic features characteristic of CSM interaction
that have been recently reported by \citep{And17} and would not, in any
obvious way, produce a transient in 1954.

\Sect{conclude} summarizes the strengths and weaknesses of the
various models and makes some suggestions for future observations.

Where stellar and supernova models are employed, they have all been
calculated using the KEPLER code \citep{Wea78,Woo02} using physics
described in \citet{Suk16}, \citet{Woo17}, and \citet{Suk18}.

\section{Circumstellar Interaction}
\lSect{csm}

Circumstellar medium (CSM) interaction is interesting both as a
generic model for iPTF14hls, and as a way to set some useful fiducial
characteristics for later use. iPTF14hls emitted at least $2.2 \times
10^{50}$ erg of light over a period of approximately 600 days
\citep{Arc17} (additional light may not have been optical), implying
an average, though variable, luminosity $\sim5 \times 10^{42}$ erg
s$^{-1}$. In the CSM interaction model, this light is emitted as the
outer layers of a supernova plow into a lower density medium at a
radius where conversion of kinetic energy to optical radiation was
efficient, i.e., $\sim10^{15}$ - 10$^{16}$ cm. An independent estimate
of the radius of the CSM comes from the $\sim600$ day duration of the
event times the highest velocity maintained throughout the event in
the spectrum, $\sim8000$ km s$^{-1}$ for H$_{\alpha}$, or $\gtaprx 5
\times 10^{16}$ cm. In order that the velocity in the spectrum not
significantly slow during the observations, there must be at least
several times more mass in the interacting ejecta than in the CSM. The
kinetic energy in the ejecta must also be at least several times what
was seen in radiation or no kinetic energy would be left over after
600 days. A kinetic energy $E \gtaprx5 \times 10^{50}$ erg is
implied. Observed speeds throughout iPTF14hls were 4000 (for iron) to
8000 km s$^{-1}$ (for hydrogen), so the kinetic energy implies a mass
for the impacting ejecta of at least one to several solar masses. The
presupernova may have ejected a much larger mass that moved more
slowly. These are the characteristics of just the ``working surface''
of the shock, and are minima for the total ejected mass and
energy. The mass of the swept up CSM would be comparable, but less to
avoid excessive deceleration.

For a solar mass to exist at $5 \times 10^{16}$ cm implies either
explosive mass loss or a steady loss rate $\gtaprx$0.01$v_7$
\Msun \ y$^{-1}$ where $v_7$ is the wind speed in units of 100 km
s$^{-1}$. This mass loss rate must persist for 100/$v_7$ years. Again
the actual mass loss could be bigger because only the inner, slowest
moving matter will interact and produce the high luminosity. Ejecting
one solar mass at 1000 km s$^{-1}$ requires 10$^{49}$ erg.

\subsection{Approximations and Models}
\lSect{csmmodels}

Some approximate scaling relations will be useful. Consider the
simplest case of an ejected shell with mass, $M_{\rm eject}$, and energy,
$E$, that encounters a previously ejected shell with mass, $M_{\rm
  CSM}$, outer radius, $R$, and negligible inner
radius. \citet{Che82a} has studied the general case in which both the
CSM and homologously coasting supernova ejecta have initial densities
that depend on a power law of the radius, $\rho_{\rm CSM} \propto
r^{-s}$ and $\rho_{\rm ejecta} \propto r^{-n}$ respectively, and gives
useful scaling relations for the radii and masses of the shocked CSM
and ejecta as a function of time.  Chevalier (1982ab) considered the
special case $s = 2$, which corresponds to a stellar wind with constant
speed and mass loss rate.  For simplicity, that simple case will be
adopted here, though the real situation may be different, especially
when the mass loss is explosive.

Then the unknowns are $M_{\rm eject}$, $E$, $n$, $M_{\rm CSM}$, and
$R$.  Actually it is the ratio $M_{\rm CSM}/R$ that matters, and not
the individual terms, so long as the radius of the shock remains
bounded by $R$. This is because for the special case $s$ = 2, $\rho
r^2$ = constant = $ q = M_{\rm CSM}/(4 \pi R)$. As Chevalier notes,
and as will later be confirmed here, the interaction region is thin,
so one can, to good approximation, assign a single radius, $r(t)$, to
the forward and reverse shocks \citep{Che82b},
\begin{equation}
r(t) \ = \ \left[\frac{2 U^n}{(n-4)(n-3) \, q)}\right]^{1/(n-2)} \,
t^{(n-3)/(n-2)}.
\lEq{roft1}
\end{equation}
Here $U$ is a constant used to normalize the density distribution in the
supernova ejecta, so that at time, $t$, and radius, $r$, $\rho_{\rm
  ejecta}(r,t) = t^{-3} (r/(t U))^{-n}$. For the special case $n = 7$,
\citet{Che82c} gives
\begin{equation}
r^{(7)} \ = \ 0.823 \, \left(\frac{E^2}{M_{\rm eject} q}\right)^{2/5} \,
t^{4/5},
\lEq{roft2}
\end{equation}
where the superscript 7 refers to the assumption $n = 7$. Normalizing to
conditions we will find appropriate for iPTF14hls, $E = 4 \times
10^{50}$ erg, $M_{\rm eject}$ = 1.0 \Msun, $M_{\rm CSM}$ = 0.4
\Msun, and $R_{\rm CSM} = 4.5 \times 10^{16}$,
\begin{equation}
r^{(7)}  \ = \ 1.4 \times 10^{16} \ t_7^{4/5} \ {\rm cm},
\lEq{roft3}
\end{equation} 
where $t_7$ is time in units of $10^7$ s.

As Chevalier notes, $n = 7$ is appropriate to Type Ia supernovae and
$n = 12$ may be a better choice for Type II supernovae occurring in
red supergiants. Then one must make a choice for $U$ in
\eq{roft1}. Based upon observations of Type II supernovae,
\citet{Che82b} suggests $U$ is a few times $10^9$. This is consistent
with the models of \citet{Woo07b}. Here $U = 2.5 \times 10^9$ is
adopted, giving
\begin{equation}
r^{(12)}  \ = \ 8.5 \times 10^{15} \, t_7^{9/10} \ {\rm cm}. 
\lEq{roft4}
\end{equation}

In fact, use of $n = 12$ is inexact for a Type II supernova,
especially as one goes deeper into the ejecta as is appropriate for
the present problem. The CSM is also unlikely to have the ideal s = 2
distribution of a wind at constant speed and mass loss rate, so either
formula would suffice. Here we will use $r^{(12)}$.

Continuing to assume a thin shell with only a single radius and speed,
the shock velocity is the derivative of the radius,
\begin{equation}
v_s^{(12)} \ = \ 7700 \, t_7^{-1/10} \ {\rm km \ s^{-1}}.
\lEq{voft}
\end{equation}
The bolometric luminosity is given by
\begin{align}
L \ &= \ 2 \pi r^2 \rho_{\rm CSM} \,v_s^3 \ = \ 2 \pi q v_s^3\\
    &= \ 0.5 M_{\rm CSM}/R_{\rm CSM}\, v_s^3\\
    &= \ 4.0 \times 10^{42} \, t_7^{-3/10} \ {\rm erg \ s^{-1}},
\lEq{loft}
\end{align}
where the superscript $n = 12$ has been omitted. The very slow evolution
of the velocity and luminosity in these equations resemble that seen
in iPTF14hls, and is a consequence of the steep power-law dependence of
the density in the supernova ejecta. For the $n = 7$ case, velocity and
luminosity would have declined a bit more steeply as $t^{-1/5}$ and
$t^{-3/5}$, respectively. As matter deeper in the ejecta interacts at
later times, one expects $n$ to decrease and the rate of decline to
increase.

To test these approximations, a 15 \Msun \ supernova with a total
explosion kinetic energy of $2.4 \times 10^{51}$ erg \citep{Woo07b}
was surrounded by a low density shell of 0.4 \Msun \ consisting of
hydrogen and helium with an outer radius of $4.5 \times 10^{16}$
cm. Within the shell, $\rho r^2$ was a constant, implying $q = 1.41
\times 10^{15}$ g cm$^{-1}$. This value for $q$ will prove a useful
constraint for successful CSM models throughout the paper.  The
presupernova star, without the artificial CSM, had a total mass of
12.79 \Msun, 8.52 \Msun \ of which was its low density hydrogenic
envelope with radius $5.65 \times 10^{13}$ cm. The unconfined 15 \Msun
\ supernova developed $6.6 \times 10^{50}$ erg in its outer 1.0 \Msun
\ of ejecta, but, including the CSM, a third of that energy was
radiated during the first 600 days, so this is close to the estimate
used in developing \eq{roft3}.

The resulting light curve is shown in \Fig{csmsn}. The event is
particularly bright during the first 100 days when iPTF14hls was not
well sampled because the underlying supernova contributes appreciably
to the CSM interaction in producing the total luminosity. This
contribution would be greatly reduced if the progenitor had been a
blue supergiant (BSG). It would also have been slightly reduced or
shortened if the hydrogen envelope were less massive or the explosion
energy smaller. After 100 days, the light curve agrees well with
\eq{loft} and with iPTF14hls for the fiducial parameters.

\Fig{csmun} shows the density and velocity evolution for this model
280 days after core collapse. It is interesting that this long bright
supernova can be powered by the interaction of only about 1 \Msun \ of
ejecta with a modest kinetic energy. Also interesting is the pile up
of the ejected matter and swept up CSM into a thin, dense shell. Half
way into the supernova, 0.67 \Msun \ has piled up in a thin shell with
a density roughly five orders of magnitude greater than the medium
into which it is moving. This shell moves with nearly uniform velocity
and justifies the assumption of a single radius for the forward and
reverse shocks made in the analytic approximation. In a
multi-dimensional calculation this shell would be unstable and spread
over a region $\delta r/r \gtaprx$ 10\%, but a large density contrast
would still persist \citep{Che14}.  The KEPLER code cannot accurately
calculate the properties of the photosphere for a thin shock wave in
matter that has recombined and is interacting in a region optically thin
to electron scattering. It is assumed throughout this paper that most
of the radiation comes out in the optical waveband. iPTF14hls was not
bright in radio or x-rays \citep{Arc17}. The flux is well determined
in the model from momentum conservation, but not its temperature. One
can speculate, however, that the photosphere lies within this dense
fast moving shell. If fully ionized, the shell would be optically
thick and the surroundings quite thin. Perhaps that accounts for the
broad hydrogen lines in the spectrum \citep{Che94}. Clearly much work
remains to be done on the radiation transport.

% The light curve shown is subenergetic by fac 2
%  2.47e50 erg overall but 1.03e50 erg of that was radiated in the 
%  first 100 days when recombination contributed.  

While the qualitative agreement with the light curve and speed of the
fastest moving hydrogen with what was observed in iPTF14hls are 
good, there are several deficiencies in this simple model. The light
curve lacks the ``flares `` seen in the observations, though an
inhomogeneous CSM could be invoked. The 4000 km s$^{-1}$ spectral
features go unexplained in the simple isotropic model, though a two
component, anisotropic explosion with slower speeds at some angles
might remedy that \citep{And17}. A large increase in the mass of the
circumstellar shell ($q$ in \eq{roft1}) would be required though to
slow the speed by a factor of two in some directions. The shell mass
assumed, 0.4 \Msun, was already substantial since the minimum radius
is already set by the duration of the event. The desired variation
might be more easily achieved by invoking an anisotropic explosion,
i.e., changing U.

The possible pre-explosive outburst in 1954 would also require an
additional explanation, but perhaps most challenging is the lack of a
clear explanation of just such a massive circumstellar shell came to
be ejected just before the supernova.

% fig 1 - circumstellar interaction
\begin{figure}
\includegraphics[width=0.48\textwidth]{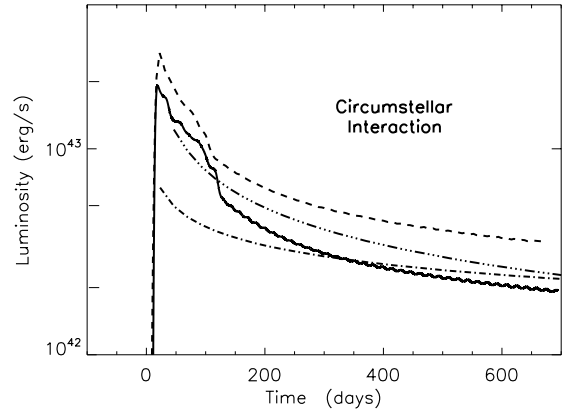}
\caption{Bolometric light curve for the explosion of a 15 \Msun
  \ supernova when that explosion is surrounded by a low density shell
  of matter with mass 0.4 \Msun \ and outer radius $4.5 \times
  10^{16}$ cm (solid dark line). The explosion energy of the supernova
  was $2.4 \times 10^{51}$ erg, but only the outer solar mass with
  initial kinetic energy $6.6 \times 10^{50}$ erg participated in the
  circumstellar interaction during the time shown. During the first
  100 days, the underlying supernova adds appreciably to the
  circumstellar interaction in producing light. Also shown as the
  dashed line is the effect of increasing the speed of the outer
  layers of the model by 25\%. The dot-dashed line shows the
  approximation discussed in the text for n = 12 (\eq{roft4}). The
  triple-dot dashed line is for n = 8, (\eq{roft3}), with $M_{\rm
    eject} = 1.0$ \Msun, $M_{\rm CSM} = 0.4$ \Msun, $R_{\rm CSM} = 4.5
  \times 10^{16}$ cm, and $E = 4 \times 10^{50}$ erg. \lFig{csmsn}}
 \end{figure}

% fig 2 - 15 + CSM interaction velocity and density
\begin{figure}
\includegraphics[width=0.45\textwidth]{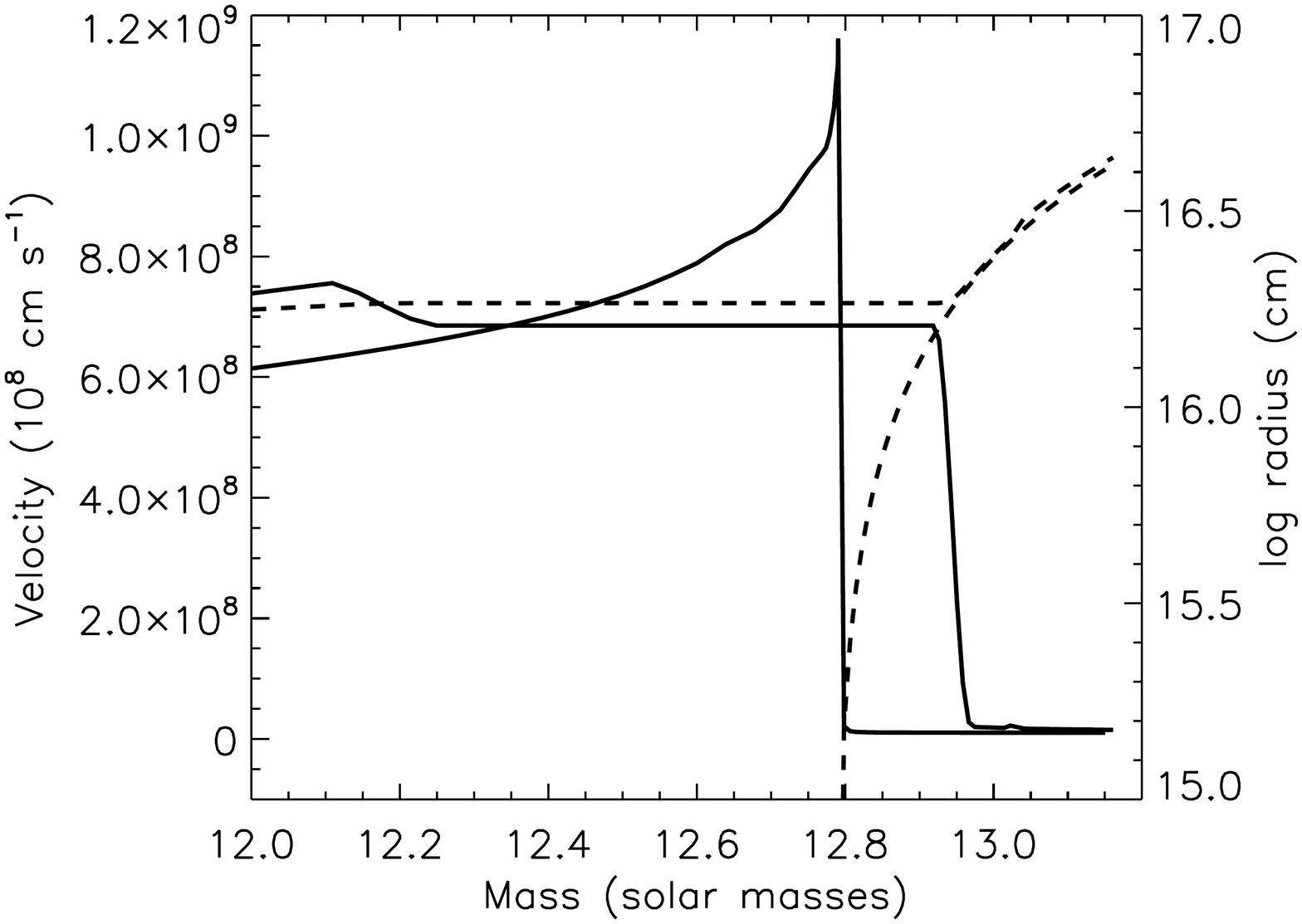}
\includegraphics[width=0.45\textwidth]{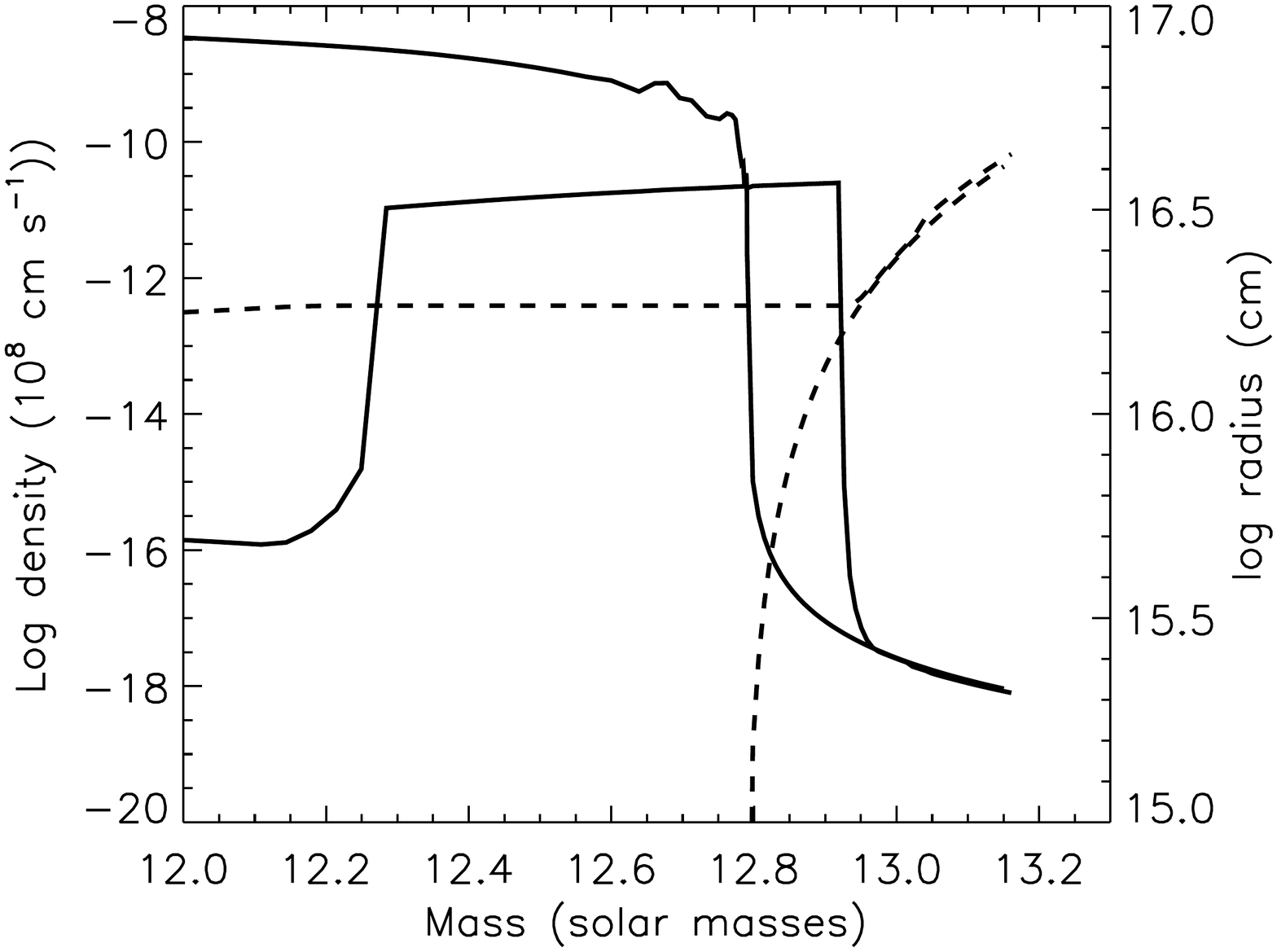}
\caption{Velocity and density for two times in the outer layers of the
  15 \Msun \ supernova model shown in \Fig{csmsn}. At the earlier time
  shown, 1.6 days after core collapse, the surface layers of the
  supernova are first encountering the circumstellar medium. The later
  time, 280 days post collapse, is about half way through the light
  curve. 0.67 \Msun \ has piled up in a thin dense shell currently at
  a radius of $1.84 \times 10^{16}$ cm moving at 6850 km
s$^{-1}$. This shell would be unstable to break up in 2 or 3
dimensions. \lFig{csmun}}
\end{figure}

\subsection{Constraints on Gravity Wave-Driven Mass Loss}
\lSect{gwave}

How did $\sim1$ \Msun \ come to reside several $\times 10^{16}$ cm
away from a dying star?  Nuclear burning time scales that affect the
stellar radius and lead to interaction in a binary system are too
long.  The time from helium depletion until carbon ignition (or
equivalently star death) is tens of thousands of years, and, for
models without gravity-wave transport, the star's radius does not
change at all during the last century.

On the other hand, a century is an inconveniently long time for
gravity wave-driven mass loss \citep{Qua12a,Qua16,Ful17,Ful17a}.
Central carbon burning can last for centuries in a common presupernova
star, but the large convective luminosities that might deliver grossly
super-Eddington powers to the hydrogen envelope develop only during
carbon shell burning. For example, a 15 \Msun \ presupernova star
\citep{Suk18} first develops a convective luminosity of 10$^{40}$ erg
s$^{-1}$ during carbon shell burning when the star has 49 years left
to live. Convective powers of 10$^{41}$, 10$^{42}$, and 10$^{43}$ erg
s$^{-1}$, in any zone, are reached only 6.9, 3.6 and 1.6 years,
respectively, before the star dies.  The regular surface luminosity of
the star during this time is $3.64 \times 10^{38}$ erg s$^{-1}$.

The vast majority of the energy developed during carbon shell burning
goes into neutrino losses. The efficiency for conversion and transport
into gravity waves that cause major changes in the envelope structure
and luminosity is uncertain, but may be $\sim0.1 - 1$\%
\citep{Ful17}. If one requires a maximum convective power of 10$^{41}$
erg s in order to compete with regular burning in the envelope, a
major augmentation to the mass loss is not expected until the the last
decade of the star's life. By then a speed of 1000 km s$^{-1}$ would be
necessary to take the matter to the necessary distance before the star
dies. This requires the delivery, during that last decade, of
$\sim10^{49}$erg s$^{-1}$ to the stellar surface, which may be more
than the model can provide unless the efficiency factor substantially
exceeds 1\%.

Moderate increases in the mass of the star do not help. For a 25
\Msun \ model, maximum convective powers of 10$^{40}$, 10$^{41}$ and
10$^{42}$ erg s$^{-1}$ are developed the last 3.4, 1.17, and 1.04
years. The star's luminosity then is $9.98 \times 10^{38}$ erg
s$^{-1}$. Certainly it is too soon to rule out gravity wave-driven
mass loss as a contributing factor in making iPTF14hls, but the
numbers are constraining.  One implication is that, if the transport
of convective power to the surface is responsible for the presupernova
mass ejection, the CSM more likely has a speed closer to 1000 km
s$^{-1}$ than 100 km s$^{-1}$. Otherwise the matter could not get to
the necessary radius in the short time during which the high power
is developed. This speed is consistent with the late time spectra
reported by \citet{And17}.

An alternative might be to use a lower mass star near 10 \Msun.
Models of these stars show, in some cases, the ejection of the entire
hydrogen envelope months to years before core collapse \citep{Woo15b}.
The driving mechanism is a degenerate silicon core flash. The
envelopes of some models reach 10$^{16}$ cm before iron-core collapse,
but the explosion energies of such low mass stars, $\sim10^{50}$ erg
may be inadequate to produce the light \citep{Suk16}.

\section{Pulsational Pair-Instability Supernovae}
\lSect{ppisn} 

PPISN occur in non-rotating stars from 80 to 140 \Msun \ that do not
lose a large fraction of their helium cores (35 to 65 \Msun) before
dying \citep{Woo07a,Woo17}. The number of pulses, their energy, and
the total duration increase with mass and range from days to
millennia. The total kinetic energy in all pulses can approach $4
\times 10^{51}$ erg, but only for the most massive models with very
long durations. For durations in the range of one to several years, as
measured from the first pulse until iron-core collapse, the initial
mass is in the range 105 to 115 \Msun \ and the total kinetic energy,
about $1 \times 10^{51}$ erg. The energy in individual pulses is
less. For durations of a century, the mass range extends to 120 \Msun
\ and up to $2 \times 10^{51}$ erg may be available. Except for the
initial pulse that ejects most of the star's envelope, the light
curves of PPISN are completely powered by colliding shells of
matter. There is no contribution from radioactivity or
recombination. PPISN light curves are thus an example of CSM
interaction.

Several possible PPISN scenarios for iPTF14hls are considered here. In
the first, the 1954 outburst \citep{Arc17} must be explained, as well
as the long event starting in 2014. In the second, the historic
outburst is left to other causes, and focus in on events whose total
duration is just a few years. In the third case, a PPISN occurs in
conjunction with an anisotropic terminal explosion generated when the
star's iron core collapses to a compact object. In all cases, the
hydrogen lines are produced by a shock impacting the inner, slowly
moving edge of the ejected envelope so those models where the envelope
was ejected many decades before, tend to be fainter.

\Tab{models} summarizes the models of all classes presented in this
paper. For the PPISN models $E_m$ is the kinetic energy of the matter
ejected in pulse $m$ and $\tau_m$ is the time in years between that
pulse and the final collapse of the iron core. For the models
considered here $m = 2$ or 3.

\begin{deluxetable*}{ccccccccccc} 
\tablecaption{Models for iPTF14hls} 
\tablehead{ \colhead{Model}               & 
            \colhead{M$_{\rm ZAMS}$}         & 
            \colhead{M$_{\rm He}$}          &
            \colhead{M$_{\rm env}$}         & 
            \colhead{E$_1$}               & 
            \colhead{E$_2$}               &
            \colhead{E$_3$}               &
            \colhead{$\tau_1$}            &
            \colhead{$\tau_2$}            &
            \colhead{$\tau_3$}            &
            \colhead{comment}             
            \\
            \colhead{}                        &
            \colhead{[\Msun]}                 & 
            \colhead{[\Msun]}                 &
            \colhead{[\Msun]}                 &
            \colhead{[10$^{50}$ erg]}          &
            \colhead{[10$^{50}$ erg]}          &
            \colhead{[10$^{50}$ erg]}          &
            \colhead{[yr]}                    &
            \colhead{[yr]}                    &
            \colhead{[yr]}                    &
            \colhead{}                        
            }\\
\startdata
\texttt{S15}   & 15 & 4.27 & 8.52 & 24 & -- & -- & -- & -- & -- &  Ordinary SNII + CSM  \\
\vspace{0.05cm}\\
\texttt{B120}  & 120 & 54.70 & 11.11 & 7.50 & 6.79 &  --  & 63.0 & 44.1 &  --  &  BSG PPISN, 2 pulse \\
\texttt{T115}  & 115 & 52.93 & 11.35 & 7.50 & 5.27 & 3.56 & 1198 & 1152 & 1152 &  RSG PPISN, long delay \\
\texttt{T115A} & 115 & 50.47 & 29.00 & 4.55 & 4.04 & 2.30 & 4.17 & 2.44 & 0.21 &  RSG PPISN, short delay \\
\texttt{T110A} & 110 & 49.68 & 18.73 & 4.72 & 1.66 & 1.22 & 12.3 & 2.31 & 2.29 &  RSG PPISN, short delay \\
\texttt{T110B} & 110 & 49.50 & 34.12 & 5.15 & 2.20 & 0.76 & 2.92 & 0.20 & 0.15 &  RSG PPISN, short delay \\
\vspace{0.05cm}\\
\texttt{20A}   & 20  & 6.17  & 9.76  & 12.1 & 0.05 &  --  &  --  &  --  &  --  & magnetar, B const \\
\texttt{20B}   & 20  & 5.83  & 1.58  & 10.7 & 0.01 &  --  &  --  &  --  &  --  & B const, low-M envel    \\
\texttt{20C}   & 20  & 5.83  & 1.58  & 10.7 & 8.9  &  --  &  --  &  --  &  --  & B decay, low-E magnetar  \\
\texttt{20D}   & 20  & 6.17  & 9.76  & 12.1 & 139  &  --  &  --  &  --  &  --  & B decay, hi-E magnetar
\enddata
\tablecomments{For the magnetar models, E$_1$ is the energy input by the piston and E$_2$ the kinetic energy 
input by the magnetar.}
\lTab{models}
\end{deluxetable*}

\subsection{Models for iPTF14hls That Could Give a Transient in 1954}
\lSect{1954}

For this class of model, pulsing activity must span many decades and
the actual death of the star actually occurs long after pulses have
ceased. The main sequence mass range is on the higher end, 115 - 120
\Msun \ and the helium core mass, 53 - 55 \Msun. A star remains at the
site of iPTF14hls, shining at approximately the Eddington luminosity,
(10$^{40}$ erg s$^{-1}$), and will continue to do so for decades to
centuries before the iron core finally collapses, possibly
uneventfully to a black hole of about 50 \Msun.

\subsubsection{A Two-Pulse Model}

The simplest PPISN model is one with only two pulses. The first ejects
most of the envelope, and the second, much later, the rest of
the envelope and part of the helium core. The mass from this second
ejection collides with the slowly moving inner edge of the first,
illuminating a bright supernova by CSM interaction.  Still later, the
remainder of the core completes silicon burning and collapses to a
black hole. 

An example is Model B120 of \citet[][\Tab{models}]{Woo17}. This star
was, by construction, a BSG derived from a 120 \Msun \ main sequence
model. It died with a residual hydrogen envelope of 11.1 \Msun,
radius, $6.1 \times 10^{12}$ cm, and helium core mass, 54.7 \Msun. The
first pulse ejected 9.8 \Msun \ of the envelope with an energy $7.5
\times 10^{50}$ erg producing a supernova. 1.3 \Msun \ of
hydrogen-rich material remained bound. Nineteen years later, a second
strong pulse ejected 5.1 \Msun \ with an energy of $6.8 \times
10^{50}$ erg. The peak speed at the leading edge of this second
ejection was 7300 km s$^{-1}$, which declined to 5000 km s$^{-1}$ 1.0
\Msun \ into the ejecta. This 1 \Msun \ of matter had a kinetic energy
of $4.0 \times 10^{50}$ erg.  The inner 0.4 \Msun \ of the ejected
envelope with which this interacted was contained within a radius of
$4 \times 10^{16}$ cm. These numbers are quite similar to the fiducial
values required in \Sect{csmmodels} to describe iPTF14hls, though a
slightly more energetic pulse 2 would be preferable. 19 years is also
too short to explain the 1954 transient. 44 years after the second
pulse, the remaining 50.9 \Msun \ core collapsed to a black hole

\Fig{b120a} and \Fig{b120} show the light curves resulting from the
two pulses. The first explosion is relatively faint, owing to the
small radius of the BSG progenitor. \citet{Arc17} reported an absolute
magnitude of $\approx -15.6$ for the 1954 outburst at the site of
iPTF14hls, but noted that this was a lower bound to the peak
luminosity. This magnitude corresponds to a luminosity of roughly $5
\times 10^{41}$ erg s$^{-1}$, in reasonable agreement with the plateau
of the model. This luminosity only lasted for about a month in the
model though, and it would have been fortuitous to have detected
it. \Fig{b120a} also shows a bright ``tail'' on the light curve after
day 50. Not only is this emission fainter than the observations in
1954 require, but it is probably also an overestimate for the
model. This late emission results from the fallback and accretion of
the innermost ejecta onto the remaining star. If the matter were fully
ionized, the Eddington limit would be near 10$^{40}$ erg s$^{-1}$. The
matter falling back in the code has recombined though, and its opacity
is low, thus the effective Eddington luminosity is high.

Another possible source of luminosity is the interaction of the
ejected envelope with mass lost before the explosion. The outer 0.1
\Msun \ of the material ejected by the first pulse moves with speeds
7000 to 12000 km s$^{-1}$and contains $7 \times 10^{49}$ erg. Taking a
shock speed of 10000 km s$^{-1}$ as typical, a presupernova mass loss
rate 10$^{-4}$ \Msun \ y$^{-1}$, and a wind speed of 100 km s$^{-1}$,
the luminosity from CSM interaction would have been $\approx
0.5 \dot M v_{shock}^3 v_{wind}^{-1} = 3 \times 10^{41}$ erg
s$^{-1}$, near the observed value, for years. It is not obvious though
that this radiation would be mostly in the optical.

The solid line in the first panel of \Fig{b120} shows that the light
curve resulting from the second pulse continues for years after
reaching a peak value of $5.3 \times 10^{42}$ erg s$^{-1}$. The
material close to the exploding core was not very finely zoned in the
calculation - a few hundredths of \Msun - and the rise time would have
been shorter in a more finely zoned model. The density distribution in
the slowly moving innermost layers of the ejected envelope scales
roughly as r$^{-1}$ to r$^{-2}$. Its velocity increases radially
outwards from a few hundred to 1000 km s$^{-1}$.  The approximation $s
= 2$ used in developing \eq{roft1} is thus roughly applicable. The
density in the outer interacting regions of pulse 2 obeys an
approximate power law $\rho \propto r^{-n}$ with $n \approx 4$, much
shallower than usually assumed for core-collapse supernova. The
shock velocity varied from 7300 km s$^{-1}$ at the onset to 5600 km
s$^{-1}$ on day 275 to 4600 km s$^{-1}$ on day 600. At those same
times the
%models from b120b0fi#68000 and 70000 and  64000
shock interaction radius moved from $7 \times 10^{15}$ cm to $2.3
\times 10^{16}$ cm to $3.5 \times 10^{16}$ cm.  The bottom panel of
\Fig{b120} shows the velocity and radius near peak light.

In two sensitivity studies, the interval between the first and second
explosions was increased to 57 years, compatible with the 60 year
interval observed for the 1954 transient, and the velocity of the
second pulse was increased. The dashed line in the top panel of
\Fig{b120} shows that the lower circumstellar density resulting from
the longer delay, by itself, makes the light curve too faint. A much
brighter light curve, more compatible with the observations,
results in this same model if the velocity in just the outer solar
mass of ejecta in pulse 2 is increased by 50\%. As expected from
\eq{loft}, allowing the ejected envelope to expand by an additional
factor of three decreases $q = M_{\rm CSM}/R_{\rm CSM}$ by three and
decreases the early luminosity by that factor. The time scale for
decline is slower though, due to the decreased density in the factor
$U$ in \eq{roft1}, so at late times the difference is less. Increasing
the velocity raises the luminosity by roughly a factor of $v_s^3$, or
3.4. A similar multiplication of the pulse speed in the standard model
with pulse interval 19 years would also have raised the solid line by
the same factor, though this is not plotted.  In the high velocity,
long interval case (triple dot dashed line in the figure) the speed of
the highest material, just inside the reverse shock was still 8000 km
s$^{-1}$ at day 600 and its radius was $5.4 \times 10^{16}$ cm. The
unmodified pulse 2 had a kinetic energy of $6.8 \times 10^{50}$ erg;
the one with the artificial velocity increase, $1.2 \times 10^{51}$
erg, which strains the limits expected for PPISN in this mass range,
but may be feasible (\Sect{conclude}).
 
In summary, simple two-pulse models like B120 can explain the 1954
transient as well as the long duration and luminosity of iPTF14hls,
but struggle to produce the bright luminosity and high $H_{\alpha}$
velocity. An artificial adjustment to the velocity can remedy the
situation, but requires doubling the energy in the outer solar mass of
pulse 2. A brighter light curve would also result if if the interval
between pulses 1 and 2 was shortened, but that would mean giving up a
PPISN solution for the 1954 transient and having a shock speed that,
especially at late times, was slower.  The models calculated here give
smooth light curves and lack the irregularity of iPTF14hls. Clumpy
ejecta could be invoked though, and might be reasonable. The two
pulses may not have been perfectly isotropic. The reverse shock in the
first pulse might have caused some mixing in the envelope. Each
``pulse'' actually includes multiple subpulses as the core ``rings''
after the explosion. It is only the innermost, slowest part of the
ejected envelope that participates in making the light curve and the
density there is sensitive to events at the ``mass cut''.  

Making the entire event with a single pulse gives no natural
explanation for the two velocity components seen in the spectrum
though matter moving at 4000 km s$^{-1}$ is certainly present. Might
models with more pulses fare better, or is there really just one shell
seen to different depths? Interestingly the 4000 km s$^{-1}$ point in
this model is located in hydrogen-deficient material, just inside the
outer edge of the former helium core.

% b120b0fi - max at cycle 64000 
%           v = 8000 
%            4 x 10**7 s later at cycle 68000 v still 6000 
%
% light curve b120b0fi not what is plotted. 
% Light curve plotted is bsg120b0
%      velocity plotted is from bsg120b0#45000, about 6e6 s
%      after light curve rose to 10**42
% Velocities are higher but L lower in fine zoned model
%  probably a lower rho r**2

% fig 3 - 120Msun BSG Pulse 1
\begin{figure}
\includegraphics[width=0.45\textwidth]{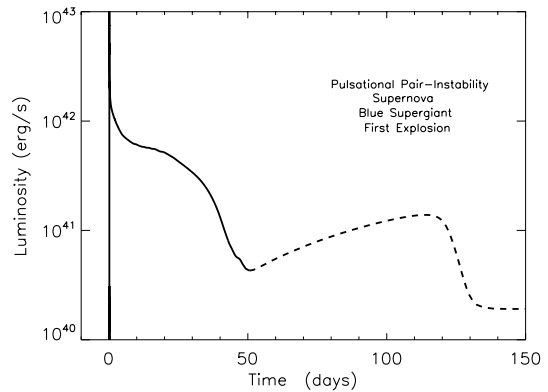}
\caption{Light curve resulting from envelope ejection caused by the
  first pulse in blue supergiant Model B120. After 50 days the light
  curve is inaccurately determined due to the inadequate treatment of
  fallback in the code. \lFig{b120a}}
\end{figure}

% fig 4 - 120Msun BSP
\begin{figure}
\includegraphics[width=0.45\textwidth]{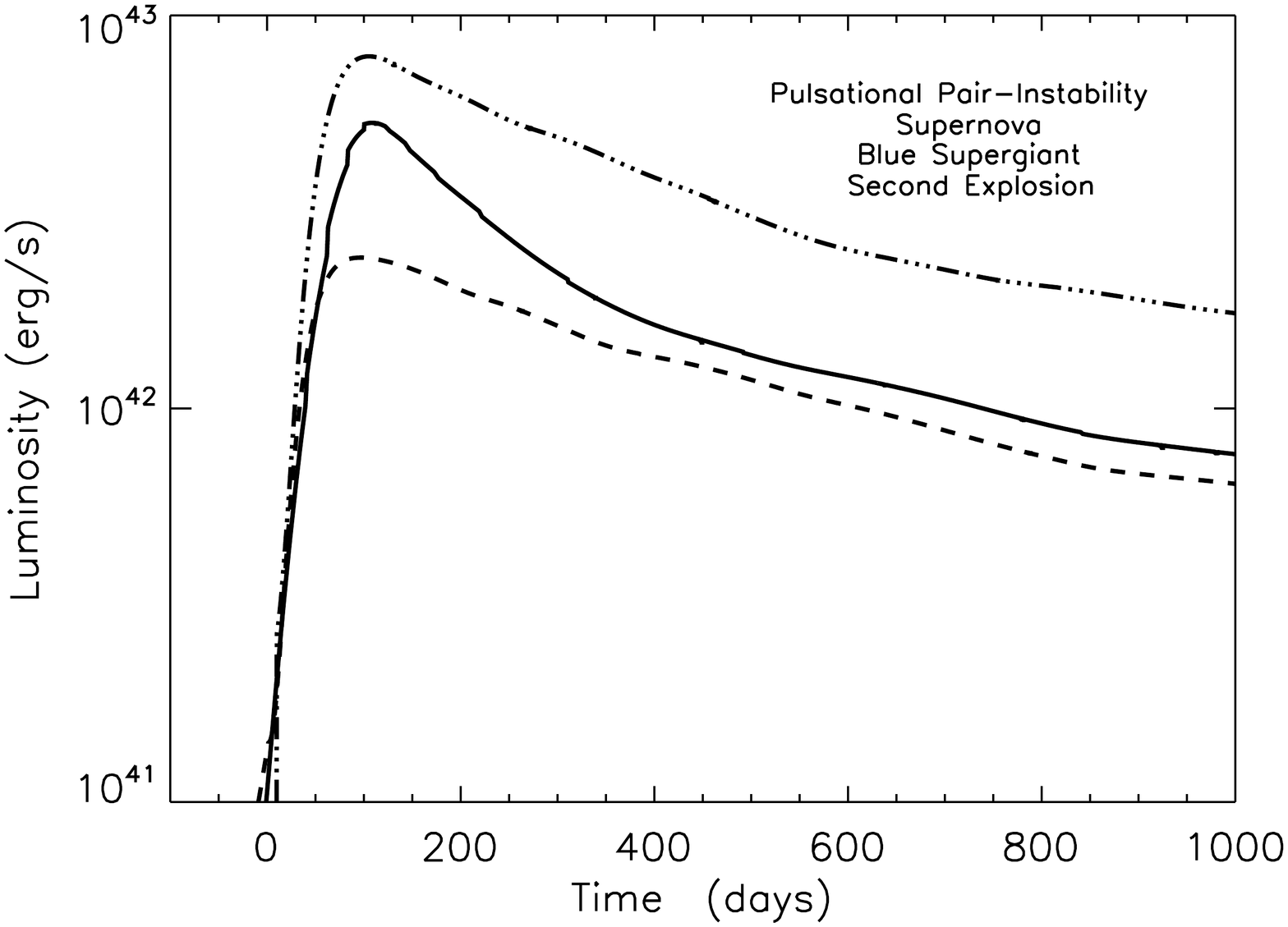}
\includegraphics[width=0.45\textwidth]{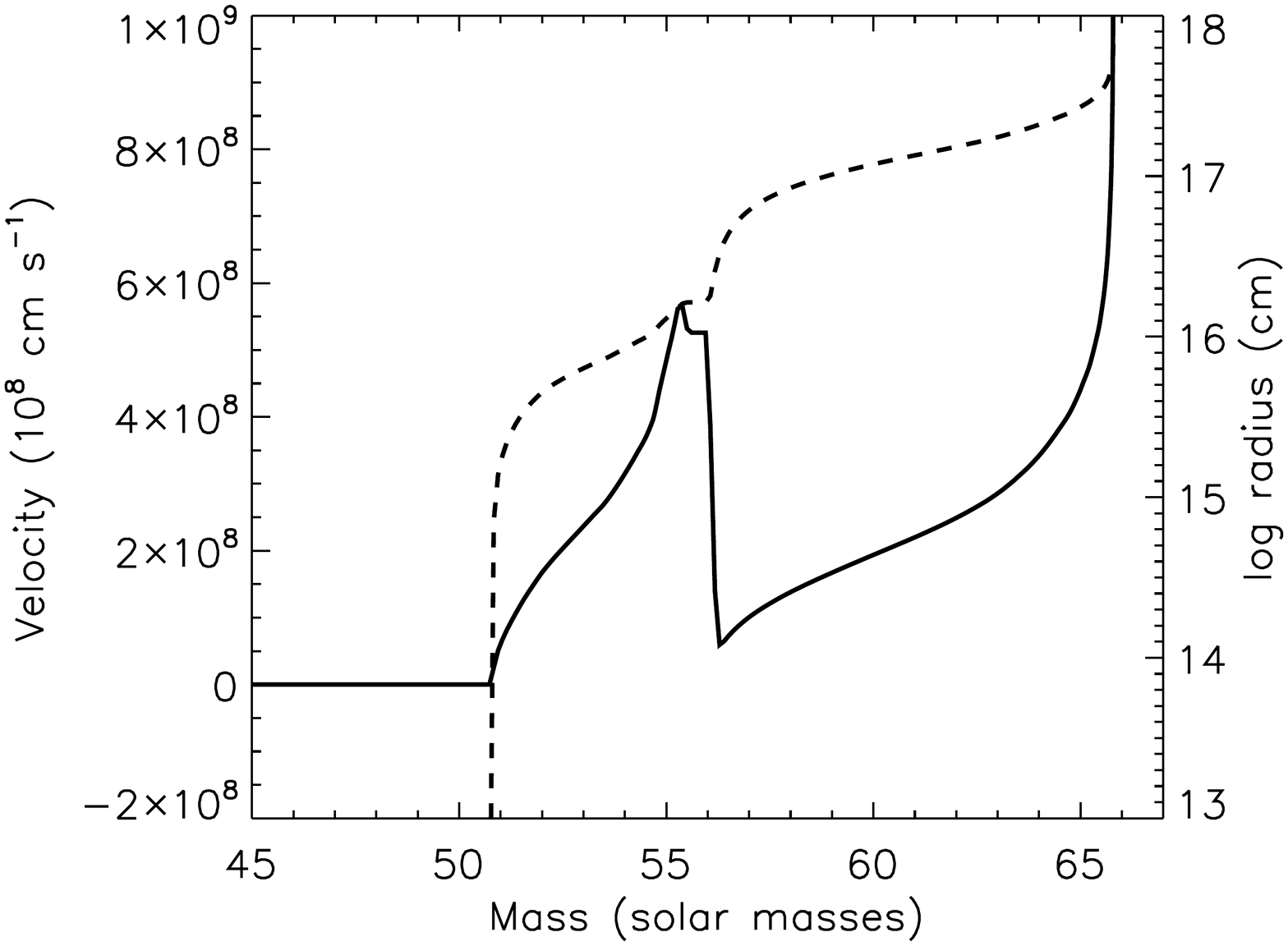}
\caption{(Top:) Light curve from the second and final pulse of Model
  B120. For the standard model (solid curve) the time is 19 years
  after the first supernova shown in \Fig{b120a}. The core of the star
  collapsed 44 years later. The dashed line shows the light curve if
  the interval between the first and second explosions is increased
  artificially to 57 years. The triple dot dashed line shows the
  result if the velocity of the second pulse in the model with the
  long delay is increased by 50\%. The rise times would have been
  shorter and the peak luminosity slightly higher for a more finely
  zoned model. (Bottom:) The velocity and radius for the standard
  model near peak emission.  \lFig{b120}}
\end{figure}

\subsubsection{A Three Pulse Model}
\lSect{t115}

Model T115 \citep{Woo17}, based on a RSG progenitor, also has a light
curve that resembles iPTF14hls, but had three pulses, the
latter two in rapid succession 46 years after the first. While the
total time between the first pulse and the final collapse of the iron
core was 1198 years, most of that time was spent in the final
contraction to stable silicon core burning during which no additional
pulses occurred.

The presupernova radius of Model T115 was $1.2 \times 10^{14}$ cm; its
luminosity, $9.8 \times 10^{39}$ erg s$^{-1}$; total mass, 64.28
\Msun; and helium core mass, 52.93, \Msun. These masses differ
slightly from those in Table 2 of \citet{Woo17} because the model was
rerun for this paper with a slightly different surface boundary
pressure and zoning. The original model had a helium core of 53.09
\Msun and pulsed for only 17 years instead of 46 years, showing the
strong sensitivity of pulse intervals to small changes in the
model. The first pulse in revised Model T115 ejected 10.4 \Msun \ of
envelope with a kinetic energy of $7.5 \times 10^{50}$ erg and a
typical speed of about 2000 km s$^{-1}$, but with a range from 0 to
7000 km s$^{-1}$. Similar to Model B120 of the previous section, 0.9
\Msun \ of envelope with hydrogen mass fraction 0.20 was not ejected
in the initial outburst. The light curve from this initial explosion
is given in \Fig{ppisnlite}.

%44.64 Msun collapsed

In this case, the first supernova was too bright to have been be the
1954 transient, unless a very substantial bolometric correction is
applied or the event was accidentally sampled during its steep
decline. After 50 days there is again a poorly determined ``tail'' on
the light curve due to fallback. The blue line in \Fig{ppisnlite}
shows the effect of reducing the radius of the progenitor to BSG-like
proportions with a radius $6.8 \times 10^{12}$ cm \citep[Model B115
  of][]{Woo17}, and the red dashed line, the effect of increasing the
opacity ($\kappa_{\rm min} = 0.01$ cm$^2$ g$^{-1}$) in the matter that
falls back. CSM interaction could again provide an enduring
luminosity, especially if a low wind speed, $\sim10$ km s$^{-1}$ is
invoked for the RSG progenitor. For a shock speed of 7000 km s$^{-1}$
and mass loss rate 10$^{-4}$ \Msun \ y$^{-1}$, the luminosity would be
$\sim10^{42}$ erg s$^{-1}$.

% fig 5 - PPISN light curves
\begin{figure}
\includegraphics[width=0.45\textwidth]{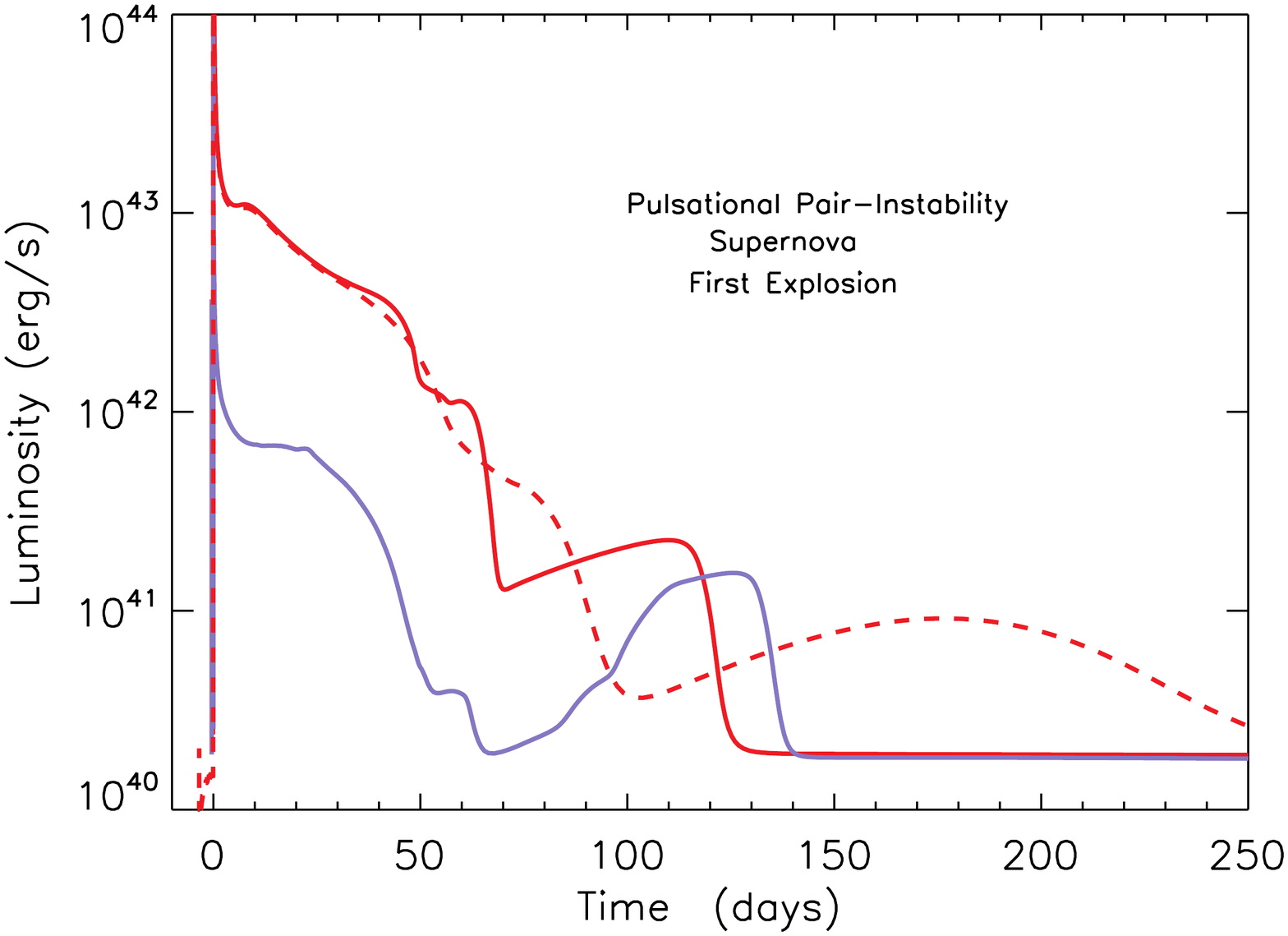}
\includegraphics[width=0.45\textwidth]{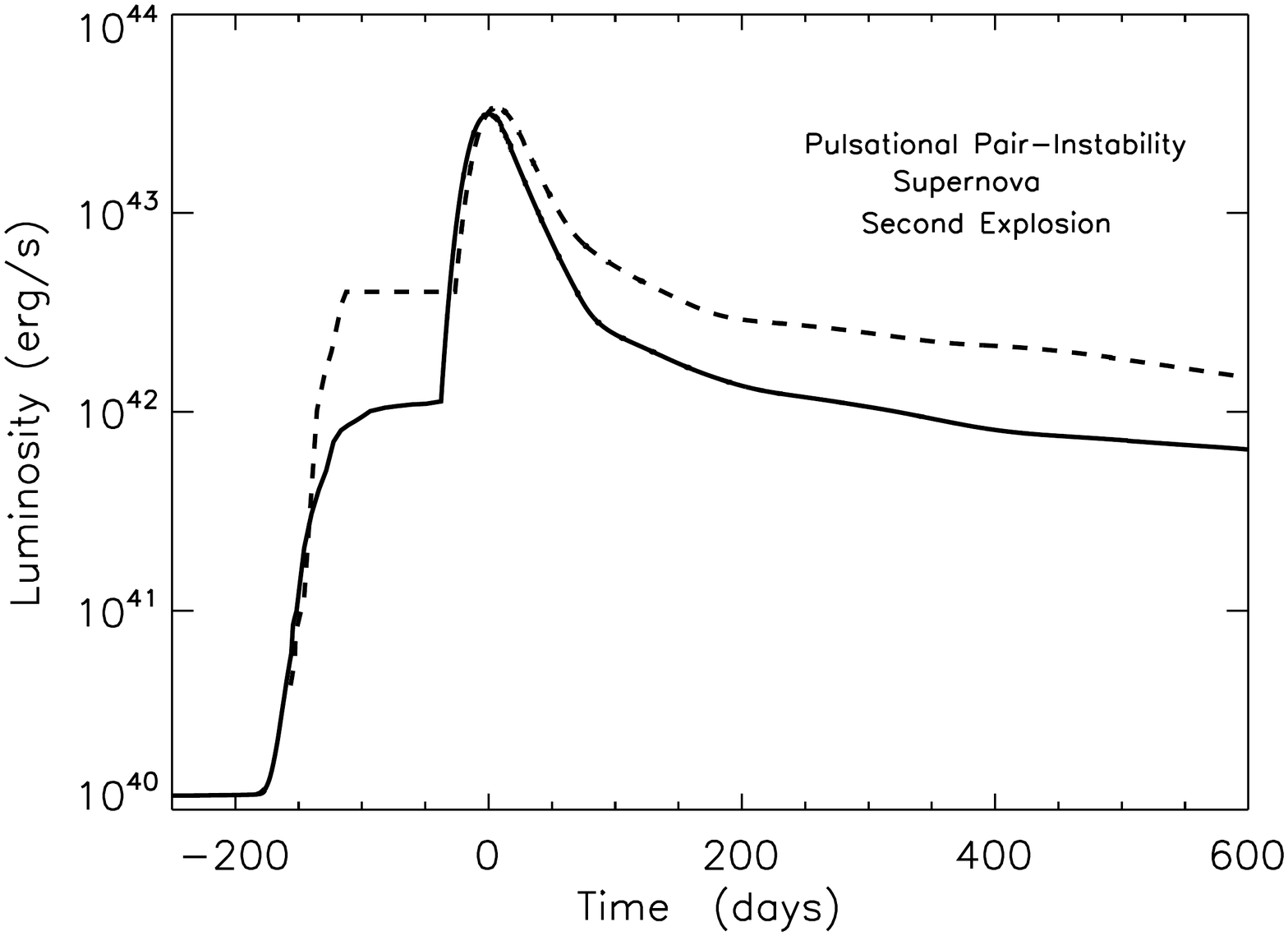}
\caption{Light curves from pulsational events in Model T115. (Top:)
  Light curve from the first pulse and ejection of most of the
  hydrogen envelope. The solid red line is for the standard red
  supergiant progenitor and the blue line is for a blue supergiant
  progenitor. The dashed red line shows the effect of using a larger
  floor to the opacity. After about 80 days most of the luminosity is
  due to fallback and accretion and is quite uncertain. (Bottom:) 46
  years later, two pulses separated by 130 days eject shells that
  collide with themselves and with the previously ejected envelope
  producing the light curve shown.  The solid curve is the standard
  model. The dashed curve results if the velocity of the leading edge
  of pulse 2 is increased by 50\% (see text).  \lFig{ppisnlite}}
\end{figure}

Forty-six years later, after ejecting most of its envelope, Model T115
experienced a second stage of thermonuclear instability during which
two additional shells of 5.6 \Msun \ and 3.5 \Msun \ were ejected in
an interval of 130 days. This added a kinetic energy of $8.9 \times
10^{50}$ erg - $5.3 \times 10^{50}$ erg in the second pulse and $3.6
\times 10^{50}$ erg in the third. The leading edge of pulse 2, located
in hydrogen-rich matter (X$_{\rm H}$ = 0.2, X$_{\rm He}$ 0.8),
initially moved at about 6500 km s$^{-1}$. Its interaction with the
slower moving material from the prior envelope ejection with speeds
$\sim300 - 500$ km s$^{-1}$ at a radius of $\sim10^{16}$ cm produced
an enduring luminosity $\sim 10^{42}$ erg s$^{-1}$ (\Fig{ppisnlite}).

These values are again in the ballpark of the analytic model
(\Sect{csmmodels}), but the velocity and CSM density are too low. The
``CSM'' comes again from the first pulse and, though 10.6 \Msun \ was
ejected, at the time of the second pulse, only 0.1 \Msun \ was within
$5 \times 10^{16}$ cm moving with a speed less than $\sim300$ km
s$^{-1}$. 0.3 \Msun \ was within $1.2 \times 10^{17}$ cm with a speed
less than 700 km s$^{-1}$. This is too little by a factor of about
three.  The density near the supernova would have been higher if the
initial pulse had less energy, if the mass of the envelope were
greater, or if the interval between pulses 1 and 2 was shortened.  The
outer 0.1 \Msun \ of shell 2 initially moved at 6700 km s$^{-1}$ and
the outer 1.0 \Msun \ had a kinetic energy of $2.7 \times 10^{50}$ erg
and an average speed of 5200 km s$^{-1}$. This is again too slow by
about 50\%.

During the interval between pulses 2 and 3, the leading edge of pulse
2 moved to a radius of $6 \times 10^{15}$ cm. Meanwhile, the third
pulse, with a leading edge speed near 4000 km s$^{-1}$, overtook the
slower moving ejecta from the second pulse, producing the bright
delayed peak in the light curve (\Fig{ppisnlite}).  The integral of
the light radiated over the period shown was $2.4 \times 10^{50}$
erg, showing the efficient conversion of the kinetic energy of the
second and third pulses into radiation. This is also the total light
observed in iPTF14hls and that is a success of the model. A better
treatment of the radiation transport is needed before anything
definitive can be said about the spectrum, but one should note the
presence, after the third pulse, of two shells bounded by two
shocks with different characteristic speeds not so different from what
was observed, 6000 km s$^{-1}$ and 4000 km s$^{-1}$. 

Reasonable adjustments to the model can bring the light curve and
hydrogen velocity more in line with observations.  Increasing the
velocity by 50\% in the outer ejecta of pulse 2, just the part moving
over 4000 km s$^{-1}$, adds $6.2 \times 10^{50}$ erg to the kinetic
energy of that pulse and gives the modified light curve in
\Fig{ppisnlite}. The shock speed at the outer edge of pulse 2 in
hydrogen-rich material now declines from 9000 km s$^{-1}$ on day 100
to 7200 km s$^{-1}$ on day 600, similar to what was observed. The
shock bounding pulse 3 decreased in speed from 4000 km s$^{-1}$ to
3000 km s$^{-1}$ during the same period, also consistent with
observations.

% fig 6 - PPISN velocity and density
\begin{figure}
\includegraphics[width=0.45\textwidth]{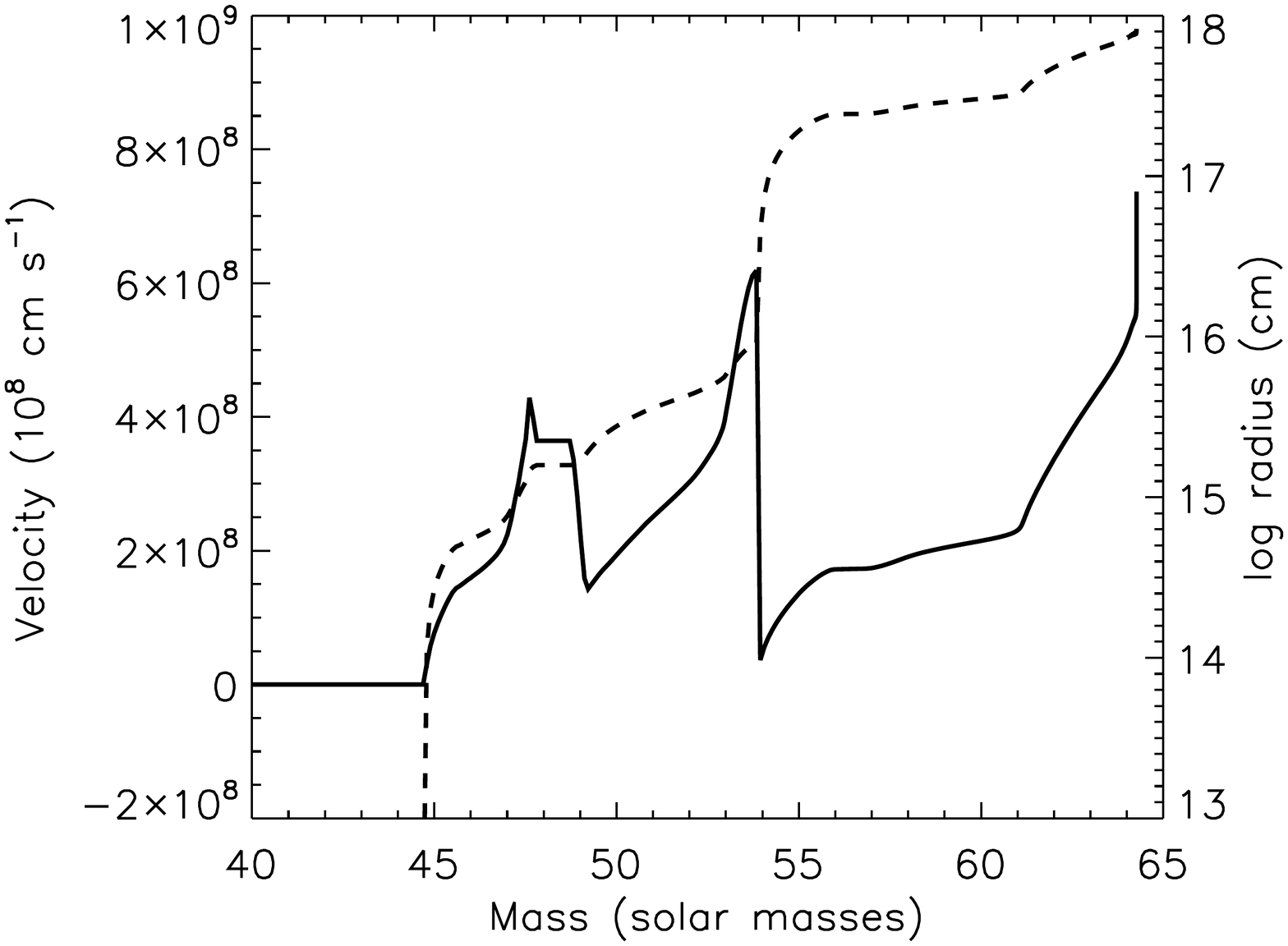}
\includegraphics[width=0.45\textwidth]{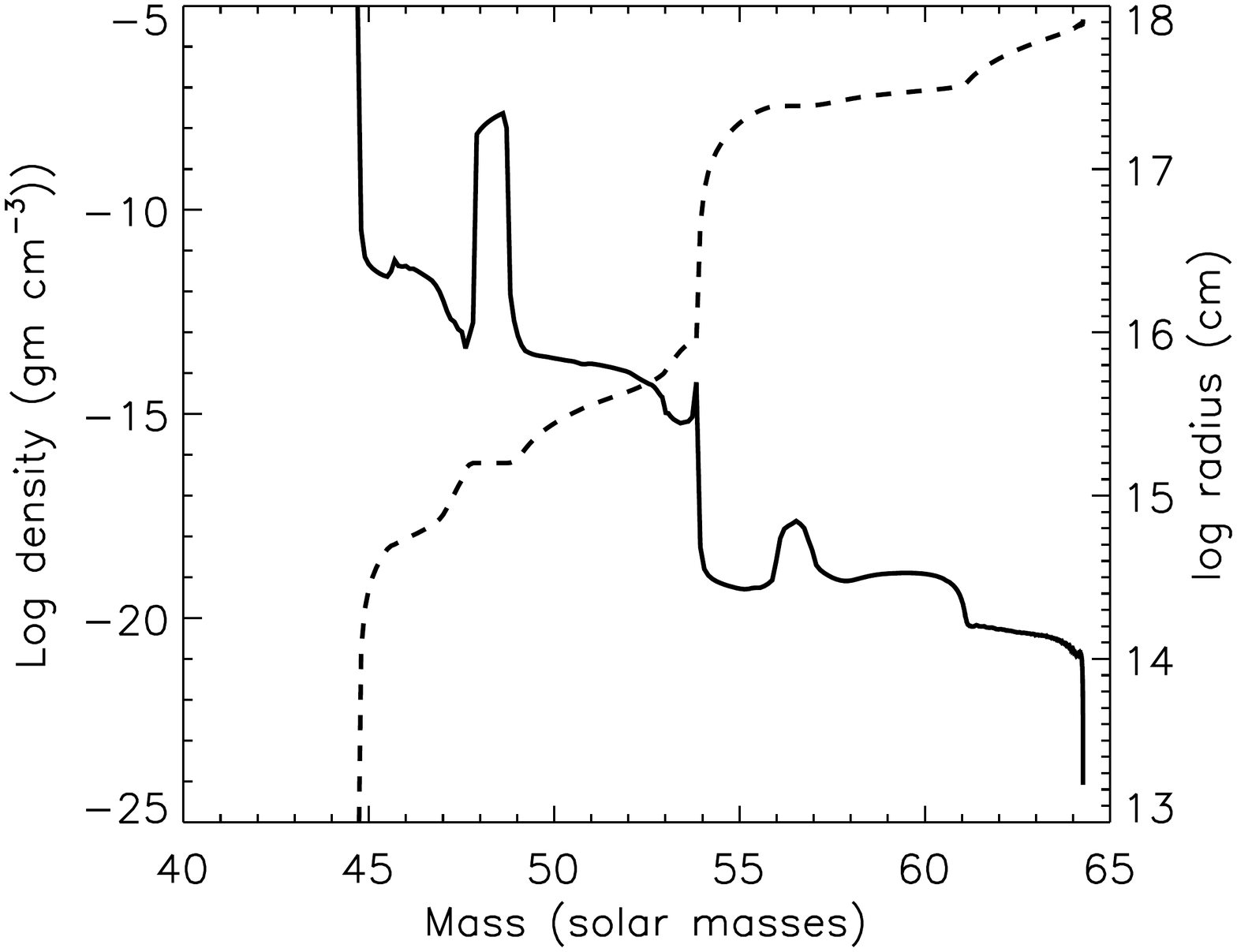}
\caption{Velocity and density near peak emission in Model T115
  (\Fig{ppisnlite}).  \lFig{ppisnun}}
\end{figure}

Like B120, Model T115 is a reasonable approximation to iPTF14hls if
one is allowed a significant modification of the energy of the second
pulse.  Several potential deficiencies remain though. The separate
4000 km s$^{-1}$ component does not appear until after the third pulse
and has a different history than the 8000 km s$^{-1}$ component. The
supernova was not spectroscopically sampled during the first 100 days
though, so perhaps this is not a problem, but the overall curve is
still too smooth. The integral under the light curve is right, but its
shape is wrong. This may reflect deficiencies in the 1D model. In
addition to the symmetry breaking conditions mentioned for Model B120,
the matter though which the shock generated by pulse 3 passes in Model
T115 has experienced mixing due to the Rayleigh-Taylor instability
\citep{Che14}. It may be clumpy and have angular and radial
variations. These would act both to broaden the peak and make the
light curve more irregular.

\subsection{Models With Shorter Delays and Prompt Black Hole Formation}
\lSect{alone}

PPISN models that produce long lasting light curves like iPTF14hls
were all previously supernovae that ejected most of their envelope in
a bright display. iPTF14hls is made by subsequent pulses running into
that ejected envelope and into each other.  Envelopes that were
ejected more recently (well after 1954) have expanded less and have a
greater $q = M_{\rm CSM}/(4 \pi R_{\rm CSM})$ in \eq{loft}.  CSM
interaction will more easily give a brighter light curve. Also the
lower mass leads to more pulses within the duration of the
light curve, giving a more irregular history resembling iPTF14hls.

Consider Model T115A of \citet[][\Tab{models} here]{Woo17}. Despite
the similarity in name, this was a different 115 \Msun \ star than
Model T115 in \Sect{t115}. Its helium core was 50.47 \Msun \ instead
of 52.93 \Msun \ and its hydrogen envelope, more massive, 29
\Msun. The initial explosion was thus more tamped and the envelope
expanded slower. The model had three pulses. The first one (actually a
pair of pulses in quick succession) ejected most of the hydrogen
envelope with a kinetic energy of $4.55 \times 10^{50}$ erg, leaving a
remnant of 53.1 \Msun, including about 2.6 \Msun \ of hydrogen
envelope.  The initial light curve (not shown) resembled the first
explosion in \Fig{ppisnlite}. 630 days later, a second strong pulse
ejected an additional 5.4 \Msun \ with kinetic energy $4.05 \times
10^{50}$ erg.  This included the residual hydrogen envelope plus the
outer edge of the helium core. 815 days after that, a third and final
pulse ejected an additional 2.0 \Msun \ of helium core with kinetic
energy $2.3 \times 10^{50}$ erg. The total kinetic energy in all three
pulses was thus $1.09 \times 10^{51}$ erg, about $4 \times 10^{50}$
erg of which was radiated away in the light curve (\Fig{longlite}).
77 days after this final pulse the star's iron core collapsed,
probably to a black hole, while the light curve was still in progress
(at 890 days in \Fig{longlite}).

% 2.06496e8   16000
% 1.51816e8   23500   630 d after pulse 1
% 8.1335e7    48200   813 d after pulse 2
% 7.4649e7 core collapse  53000  77 days after 3rd pulse
%  45.6 Msun collapsed

The initial rise in \Fig{longlite} is due the second pulse
encountering the inner edge of ejected envelope at a radius of $\sim
10^{15}$ cm. The sharp peak about 200 days later is not a new pulse,
but the same pulse encountering a thin shell of material piled up by
the reverse shock from the first pulse. In reality, this shell would
have mixed and not be so thin. The peak at 200 days would be broader,
but contain the same radiated energy.  The third and final pulse
happened at day 815 on the plot, and an additional spike is generated
at day 1490 when the piled up shell from pulse 2 running into pulse 1
is encountered by pulse 3. At this point, the core has already
collapsed and, baring additional activity generated by compact object
formation, no more mass ejection occurs.

The rapid variability in this light curve resembles that of iPTF14hls,
though it is perhaps {\sl too} variable and lasts too long. Mixing in
the ejecta would greatly reduce the variability
\citep{Che14}. Smoothing by $\delta t/t \gtaprx$ 10\% is reasonable,
and would improve the agreement with observations. Shortening the
interval between pulses 2 and 3 by a factor of two also leads to a
light curve more like iPTF14hls (the dashed line in
\Fig{longlite}). This is also a reasonable adjustment, given that the
core temperature following pulses, to which neutrino cooling is very
sensitive, may not be precisely determined.  Typical speeds in the
colliding shells are 2000 - 4000 km s$^{-1}$ though. This might be
adequate to explain the ``slow'' component seen in the iron lines of
iPTF14hls, but no hydrogen moved faster than 5000 km s$^{-1}$. Despite
many attractive features, baring some additional input of energy
(e.g., \Sect{hybrid}), this model is ruled out by the lack of high
velocity hydrogen.

% fig 7 - long lasting light curve from t115a0
\begin{figure}
\includegraphics[width=0.45\textwidth]{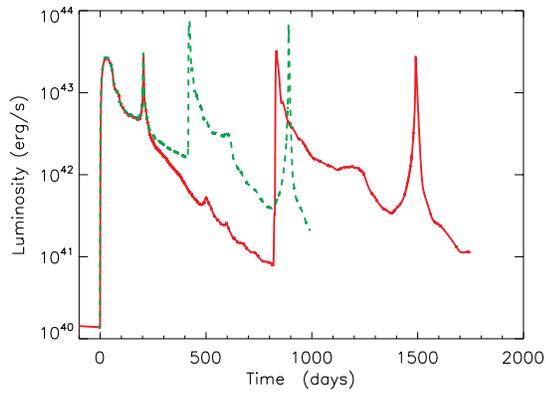}
\caption{Light curve for another 115 \Msun \ model (T115A of
  \citet{Woo17}) with a larger envelope envelope mass and slightly
  smaller helium core. The unmodified model (solid red line)
  experiences repeated outbursts over a 5 year period. The peaks at 0
  and 810 days correspond to the second and third pulses in the
  model. Spikes at 200 and 1490 days are result from the interaction
  of these pulses with thin shells of previously ejected matter (see
  text). The iron core collapsed on day 890.  The dashed green line
  shows the result if the interval between pulses 2 and 3 is
  artificially decreased by a factor of two by increasing the core
  neutrino losses. Sharp spikes in the light curve would be broadened
  by at least 100 days by mixing in a two-dimensional
  calculation. While long lasting, the hydrogenic shock moved too
  slowly in these models to be iPTF14hls. \lFig{longlite}}
\end{figure}

Model T110A was similar, but because of its lower helium core mass,
the pulses occurred in more rapid secession, resulting in a light
curve that was more continuous. The helium core mass was 49.7 \Msun,
surrounded by an envelope of 20 \Msun. A first pulse ($4.72
\times 10^{50}$ erg) ejected the hydrogen envelope 10 years prior to a
final two pulses that, in rapid succession, ejected an
additional 5.2 \Msun \ of mostly helium with an additional $2.88
\times 10^{50}$ erg. Collision of the ejected matter with the inner
edge of the previously ejected envelope, which initially had a speed
of only a few hundred km s$^{-1}$, gave the fainter light curve in the
first panel of \Fig{t110}. The model glowed continuously with
supernova-like luminosity for over 1000 days. Typical interaction
radii were 0.6 to $2 \times 10^{16}$ cm. Over the course of the light
curve, the shock speed declined from 4000 km s$^{-1}$ to 1600 km
s$^{-1}$. Most of the time it was near 2000 km s$^{-1}$. The velocity
of the hydrogen just outside the shock was $\sim$1000 km s$^{-1}$.

This shock speed is far too slow, and the light curve too faint
to be iPTF14hls. A brighter, shorter (but still 700 days long) light
curve results if the velocity of the matter ejected by this second set
of pulses is increased by a factor of 1.5 corresponding to an energy
increase of $3.5 \times 10^{50}$ erg, well within reach of the PPISN
model. Now the shock speeds are 6000 to 3200 km s$^{-1}$, but the
higher speed only lasts a short time and may not have been observed.

Model T110A is not unique. The second panel in \Fig{t110} shows that
Model T110B, with essentially the same helium core mass, 49.50 \Msun,
but larger envelope mass, 34 \Msun, has a similar light
curve. Apparently when stars with the right helium core mass, roughly
50 to 55 \Msun \ for the physics in the KEPLER code, die in stars with
substantial envelopes, they frequently make supernovae that resemble
iPTF14hls.
 
% fig 8 - 110Msun PPISN
\begin{figure}
\includegraphics[width=0.45\textwidth]{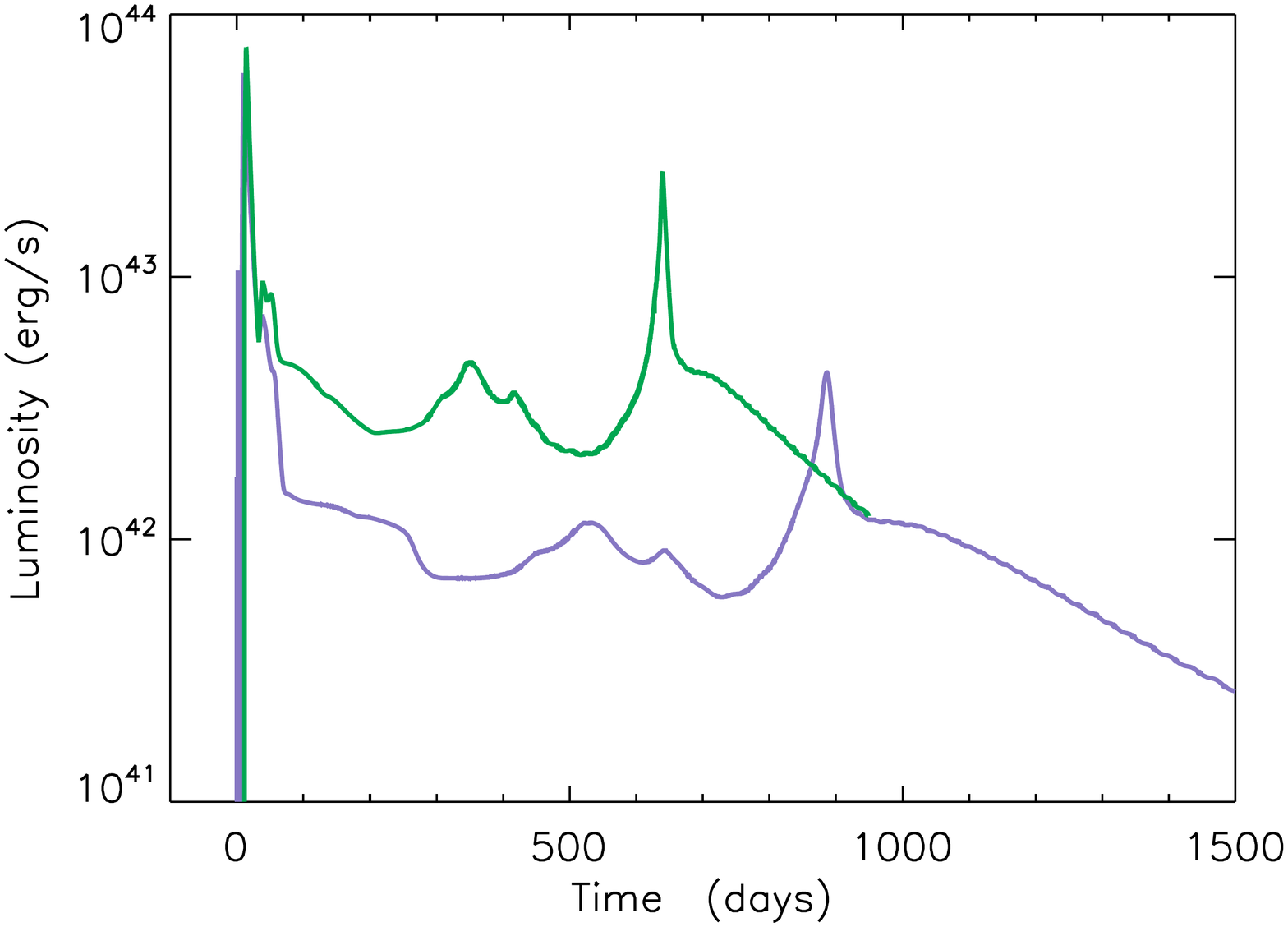}
\includegraphics[width=0.45\textwidth]{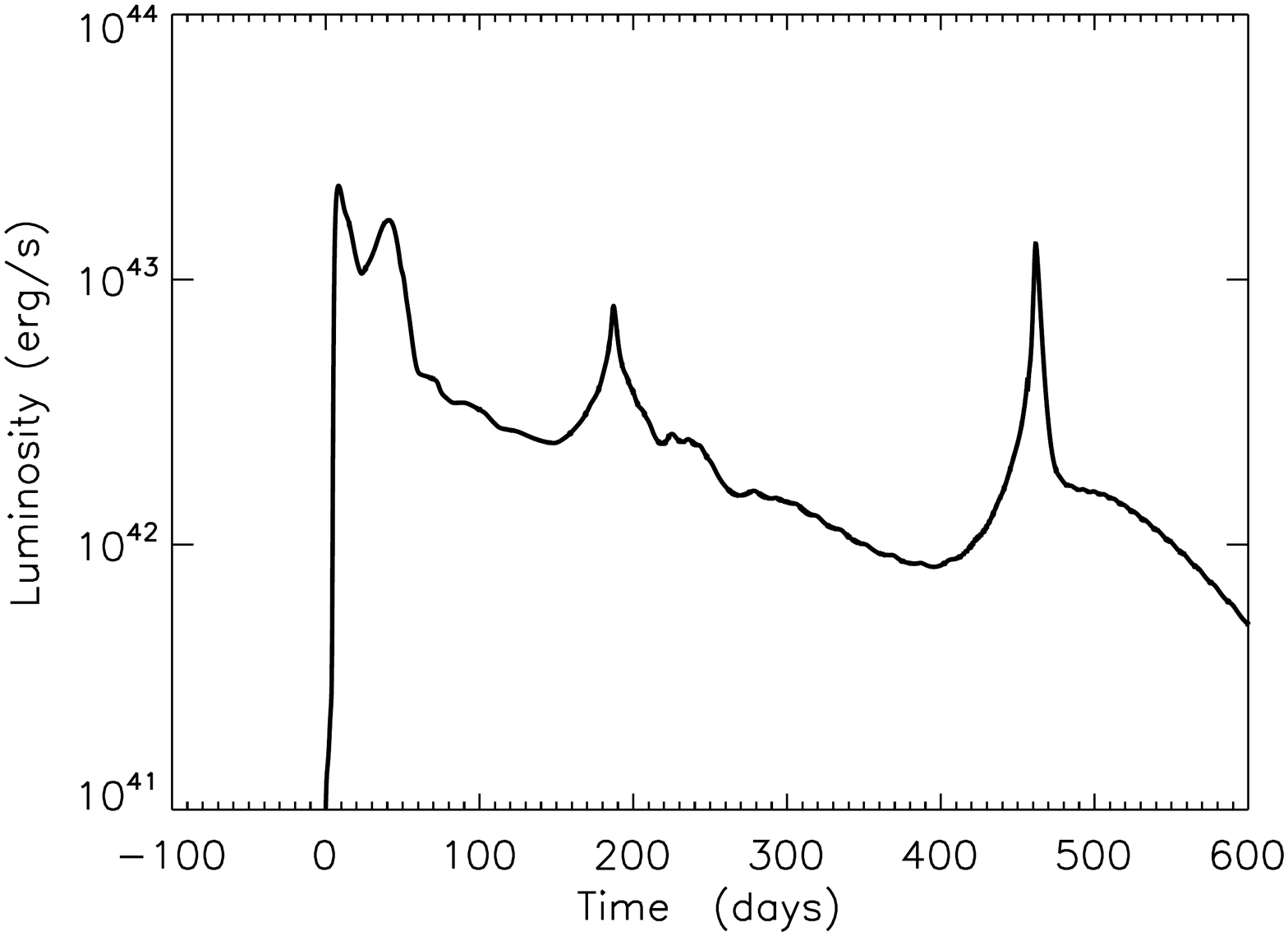}
\caption{{Top:} Explosions of 110 \Msun \ PPISN corresponding to
  Models T110A (top frame) and T110B (bottom frame) of \citet{Woo17}
  In the top frame the green curve results if the velocity is
  multiplied by 1.5.\lFig{t110}}
\end{figure}

In summary, lighter models that do not attempt to explain the 1954
transient also give light curves that, with minor adjustments, agree
with the brilliance, duration, and variability of iPTF14hls. They can
also explain the time history seen in the iron lines. Unfortunately,
all models examined so far fail to give the high velocity seen
throughout the event for the Balmer lines of hydrogen. In this regard,
they do even less well than the longer duration, more energetic events
that might also explain the 1954 transient (\Sect{1954}).

\subsection{Anisotropic Models with Terminal Explosions}
\lSect{hybrid}

All PPISN models thus far have been one-dimensional and each has
assumed the formation of an inert black hole once pulsational activity
ceases.  With the additional freedom of angular dependence and the
added energy of a terminal explosion, one can construct a broader
range of models, invoking disparate conditions at different angles for
the low and high velocity components. It becomes easier to make high
velocity hydrogen.

The chief uncertainty here is how an energetic, anisotropic flow would
develop in a thermonuclear model that is inherently
isotropic. Rotation is the obvious explanation.  There is adequate
angular momentum in some PPISN models for the iron core to become a
millisecond magnetar, though such models must avoid ever becoming
supergiants \citep{Woo17}. It would be very hard though to reverse the
inflow of collapse and eject everything outside the neutron star. One
would also expect 40 \Msun \ of oxygen and heavy elements to
eventually make their presence known. For the time being, if a
terminal, anisotropic explosion is to occur in a PPISN, it seems more
natural to invoke black hole formation. The black hole would have a
mass of about 45 \Msun \ and could be rotating rapidly with a Kerr
parameter $\sim0.1$. Polar outflows could develop from the accretion
of even a small amount of matter \citep{Mac01,Qua12,Woo12,Dex13},
provided the necessary magnetic fields could be generated near the
event horizon of the rapidly rotating hole. Accreting 0.1 \Msun \ with
1\% efficiency for conversion of rest mass into outflow could power a
10$^{51}$ erg outflow. It is interesting that this would make
iPTF14hls a close relative of gamma-ray bursts, with similar energy,
but greater baryon loading and perhaps less collimation. The main
difference here, aside from the large black hole mass, is the presence
of solar-mass shells of matter with which the accretion energized
outflow can interact.

Besides the uncertain physics of such a terminal explosion, there is
the issue of its timing. For two components to appear in the spectrum
of the same event, they must commence close together. Iron-core
collapse needs to follow swiftly on the heels of the final
pulses. There are PPISN where this is the case, though there usually
at least a few week's delay as the star goes through a final stage of
stable silicon shell burning (\Tab{models}). There is also an issue of
how to input the energy from a terminal explosion into the code. If it
comes from polar outflows driven by accretion on a massive black hole,
the energy might be in the form of a small mass moving at
semi-relativistic speed. Retaining the kinetic energy of this small
mass and not promptly radiating it away as something resembling a
gamma-ray burst afterglow requires that the shells with which it
interacts still be optically thick. 

Consider Model T110A (\Fig{t110};\Sect{alone}) that ejected most of
its hydrogen envelope 12 years before two final pulses and the
collapse of its iron core (\Tab{models}). Unfortunately in that model
iron-core collapse did not occur until 830 days after the last
pulse. Adding a high velocity component in just the final few hundred
days would not explain the observations.  In Model T110B, the delay
between the final pulse and core collapse was just 55 days. The
difference was a slightly higher central temperature, $1.36 \times
10^9$ K versus $1.13 \times 10^9$ K immediately after the pulses. A
high velocity component in Model T110B, however, gave too bright and
too brief a light curve because the hydrogen envelope had not expanded
enough. The first supernova was too close to core collapse.  Model
T115B \citep{Woo17} had a strong final pulse just 12 days before core
collapse. It's inner solar mass had already turned to iron after that
pulse, so silicon shell burning was of limited duration.

In order to show what might happen in a model capable of producing
comparable luminosities in the large angle and polar outflows,
attention was focused on Model T110A. Energetic explosions were
introduced just 10 days after the final pulse \Fig{t110}. These were not
explosions of the core though. The remaining 44.6 \Msun \ core was
excised, presumably to make the black hole, and 0.1 \Msun \ of the
matter in the inner edge of the last shell to be ejected given a high
speed corresponding to a energies of 24, 12, and 6 $\times 10^{51}$
erg. Since this was a one-dimensional simulation, these are the
equivalent isotropic energies, that are really assumed to pertain to
just some small solid angle, assumed here to be 10\% of the sky. So
the 12 $\times 10^{51}$ erg case, for example, only requires an energy
input of $1.2 \times 10^{51}$ erg.

The resulting light curves are shown in \Fig{hyper}. They are very
bright. Just 10\% of the luminosity of the 12 $\times 10^{51}$ erg
model could power the observed light curve of iPTF14hls and then
some. The velocity history (\Fig{hyper}) is also in reasonable
agreement with what was observed for the H$_{\alpha}$ line, especially
if the event was not observed in the first 100 days. The small mass of
hyper-velocity ejecta impacts the two shells ejected by the final two
pulses and sweeps them up. Because the collision occurs in a region
that is still marginally optically thick, kinetic energy is conserved
and not promptly radiated away.  Models in which the energy was
injected at day 60 fared less well and only the most energetic case
retained enough kinetic energy after the break out transient to power
a light curve like iPTF14hls - again assuming a solid angle of 10\% of
the sky.

In summary, a two component PPISN model with a terminal explosion is
contrived, but could potentially explain the major features of
iPTF14hls. The model requires both the production of a mildly
relativistic jet, for which a physical basis is lacking, and that
the core collapse very shortly after the last pulse of the PPISN. A
similar model was proposed by \citet{Woo17} to explain superluminous
supernovae, but there the timing of the core collapse was not so
constrained by the need to produce a long light curve with constant
velocity components. Also, in the models of \citet{Woo17}, the entire
helium and heavy element core within the given solid angle was
ejected, not just a small mass outside a black hole.

% fig 9 - 110Msun jet
\begin{figure}
\includegraphics[width=0.45\textwidth]{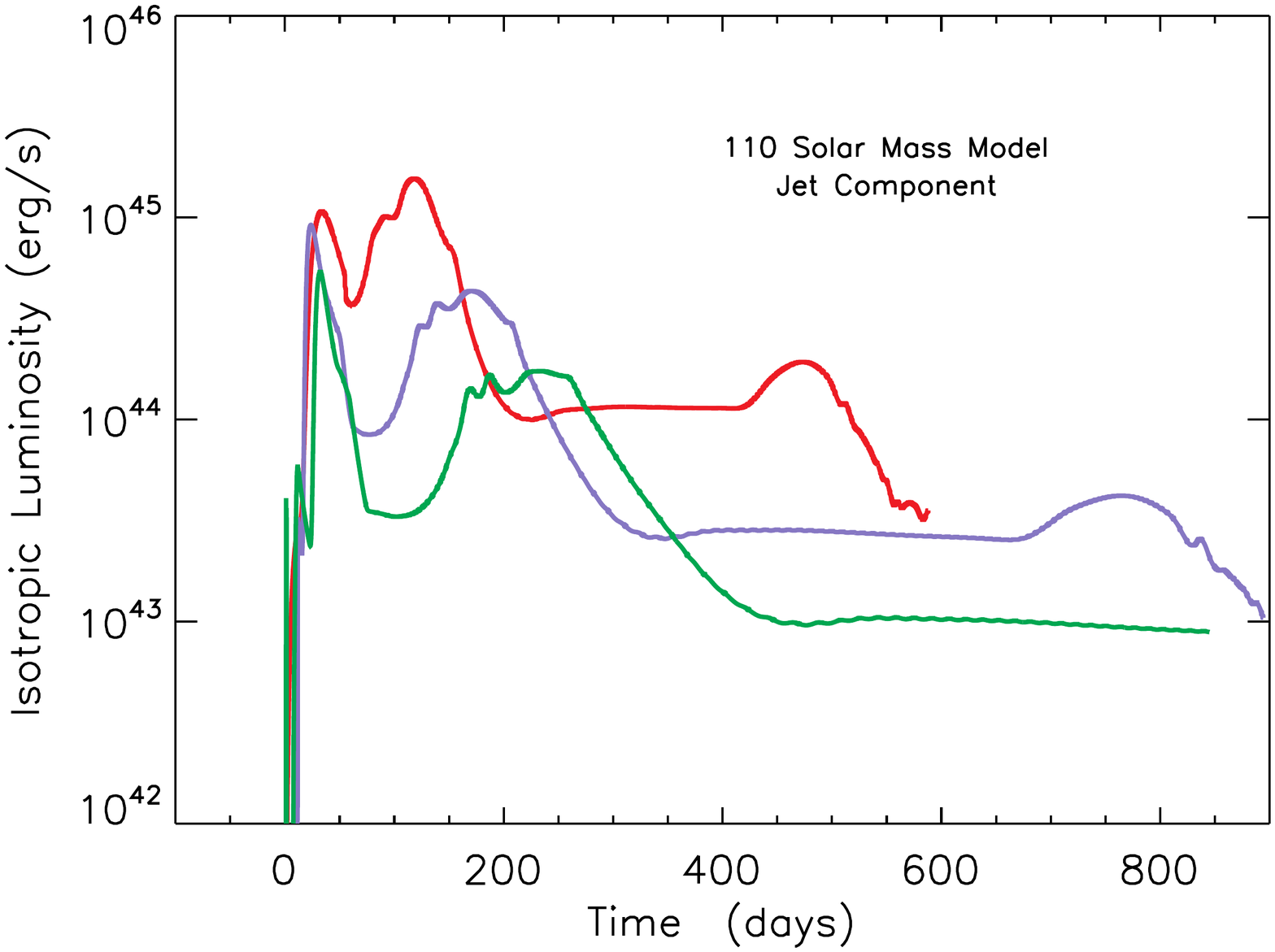}
\includegraphics[width=0.45\textwidth]{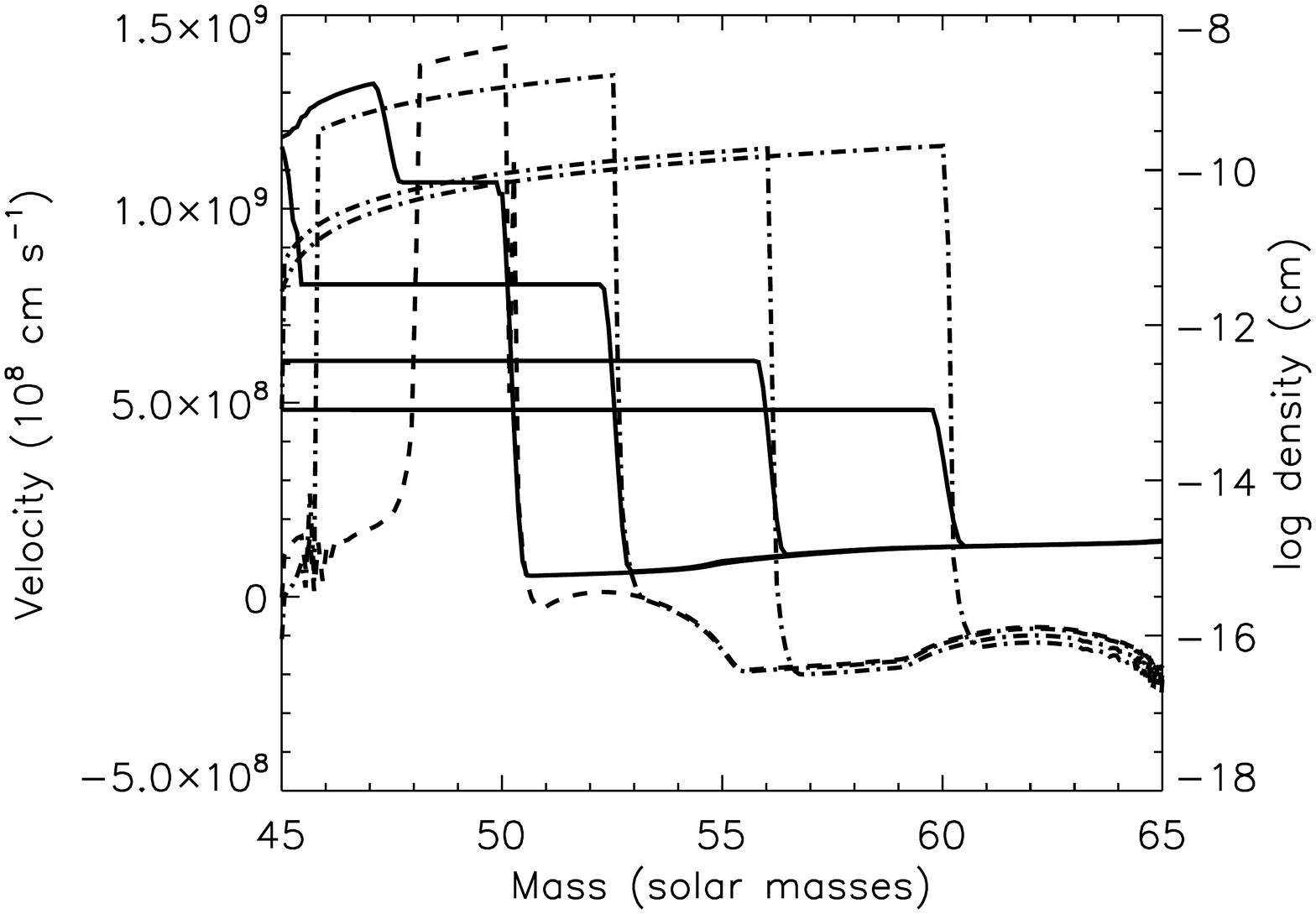}
\caption{{Top:} Light curves of three very energetic explosions
  perhaps powered by black hole accretion. The explosions were
  initiated by giving a small amount of mass a very high speed and
  allowing it to impact on the shells ejected in Model T110A (see
  \Fig{t110}). The explosions had equivalent isotropic energies of 24,
  12, and 6 $\times 10^{51}$ erg. The amount of energy radiated as
  light in the three models was 15, 7.2, and 3.6 $\times 10^{51}$
  erg. These energies and the luminosity in the figure should be
  multiplied by the fraction of the sky subtended by the mildly
  relativistic outflow, perhaps $\sim10$\%, and added to the
  luminosity for Model T110A in \Fig{t110}. (Bottom:) Velocity and
  density in the $7.2 \times 10^{51}$ erg explosion (blue line in top
  panel) evaluated at 120, 180, 440, and 730 days on the light curve
  plot.  \lFig{hyper}}
\end{figure}

\section{Magnetar Models}
\lSect{magnetar}

Magnetars are neutron stars with unusually strong magnetic fields
compared with radio pulsars. Typically a magnetar has a (dipole) field
strength $\gtaprx$10$^{14}$ G. The strong magnetic field of the
neutron star, in theory, is a consequence of very rapid rotation at
the time of its birth \citep{Dun92}. The existence of magnetars and
their key role in explaining soft gamma-ray repeaters and anomalous
x-ray pulsars is beyond doubt \citep{Kas17}. The time during which the
magnetar shines brightly is short compared with pulsars and, given
their numbers and spatial distribution, more than 10\% of all neutron
stars are probably born with these strong fields and, presumably,
rapid rotation. Rapidly rotating magnetars are also the central engine
in a leading model for gamma-ray bursts \citep[e.g.][]{Uso92} where
they are required to have magnetic fields greater than 10$^{15}$ G and
powers $\gtaprx10^{50}$ erg s$^{-1}$. Somewhere between these extremes
of rotation rate and field strength - ordinary pulsars and
ultra-powerful magnetars, $B \sim 10^{14}$ G and P $\sim$ few ms
should exist. For such characteristics, a light curve like iPTF14hls
is a natural consequence.

For lack of a deeper understanding, the magnetar energy and power of a
magnetar just after its birth are generally approximated using the
same two-parameter equations as employed for much older pulsars:
\begin{equation}
\begin{split}
E &= \frac{1}{2} I \omega^2 \\
&\approx 2 \times 10^{52} {\rm P}_{\rm ms}^{-2} \ {\rm erg}.
\end{split}
\lEq{pulse}
\end{equation}
where $E$ is the rotational energy, $I$, the moment of inertia
($\approx10^{45}$ g cm$^2$), and $P_{\rm ms}$, the period in ms.  The
approximate energy loss for dipole radiation is given by the Larmor
formula \citep[e.g.,][]{Lan80},
\begin{equation}
\begin{split}
\frac{d E}{d t} &= \frac{2}{3 c^3} \left(B R^3 \ {\rm Sin} \, \alpha \right)^2
\left(\frac{2 \pi}{{\rm P}}\right)^4 \\
&\approx 10^{49} B_{15}^2 {\rm P}_{\rm ms}^{-4} \ {\rm erg \ s^{-1}}.
\end{split}
\lEq{dipole}
\end{equation}
Here $B_{15}$ is the surface dipole field in 10$^{15}$ Gauss, $R
\approx 10^6$ cm is the neutron star radius, and $\alpha$ is the
inclination angle between the magnetic and rotational axes, taken
arbitrarily to be 30 degrees. These equation may be integrated to give
the energy and power at time $t$ \citep{Woo10},
\begin{equation}
\begin{split}
\frac{d E}{d t} &= \left(\frac{d E}{d t}\right)_0 \left(1 \, + \, t/t_p\right)^{-2} \\
E &= E_o \\
t_p &= 2000 \, P_{ms0}^2 B_{15}^{-2} \ {\rm s} 
\end{split}
\lEq{power}
\end{equation}
%green line s20D or d/magnetar1 0.6 B/4e13 always low M envelope
%blue  line s20G/magnetar1 2 B/2e15 G for 10**4 s then 0.6/4w13 low M env
%brown  s20C/magnetar2 15 B/2e15 G for 10**4 s then 0.6/4e13 high M 
%red  s20B/magnetar .6/.04  always  high M
Similar equations have been given by \citet{Kas10} with a different
choice of inclination angle. Their equations are recovered if $B$ in
the above equations is divided by $\sqrt 2$.

These equations have the simplicity of a physical model that can be
adjusted using the two parameters to fit almost any smooth light curve
so long as the emitted radiation and wind is thermalized inside the
expanding star and emitted chiefly in the optical. They are too simple
however, especially at very early times, when the neutron star and its
crust are rapidly evolving. The same magnetic field and rotation
needed to eject the accreting matter and power the supernova, a power
of at least 10$^{50}$ erg s$^{-1}$ (more if appreciable $^{56}$Ni is
to be synthesized), would rapidly sap the rotational energy and leave
the magnetar powerless during the months needed to power the light
curve. Thus we expect that $B$ in the above equation is not a constant
at early times.

Given that iPTF14hls was already well underway when discovered, one
must also include an element of uncertainty as to just when the core
collapsed and started the event. The magnetar may have deposited
appreciable energy before the supernova was discovered.

%green line s20D or d/magnetar1 0.6 B/4e13 always low M envelope
%blue  line s20G/magnetar1 2 B/2e15 G for 10**4 s then 0.6/4w13 low M env
%brown  s20C/magnetar2 15 B/2e15 G for 10**4 s then 0.6/4e13 high M 
%red  s20B/magnetar .6/.04  always  high M

% fig 10 - magnetar light curves
\begin{figure}
\begin{center}
\includegraphics[width=0.45\textwidth]{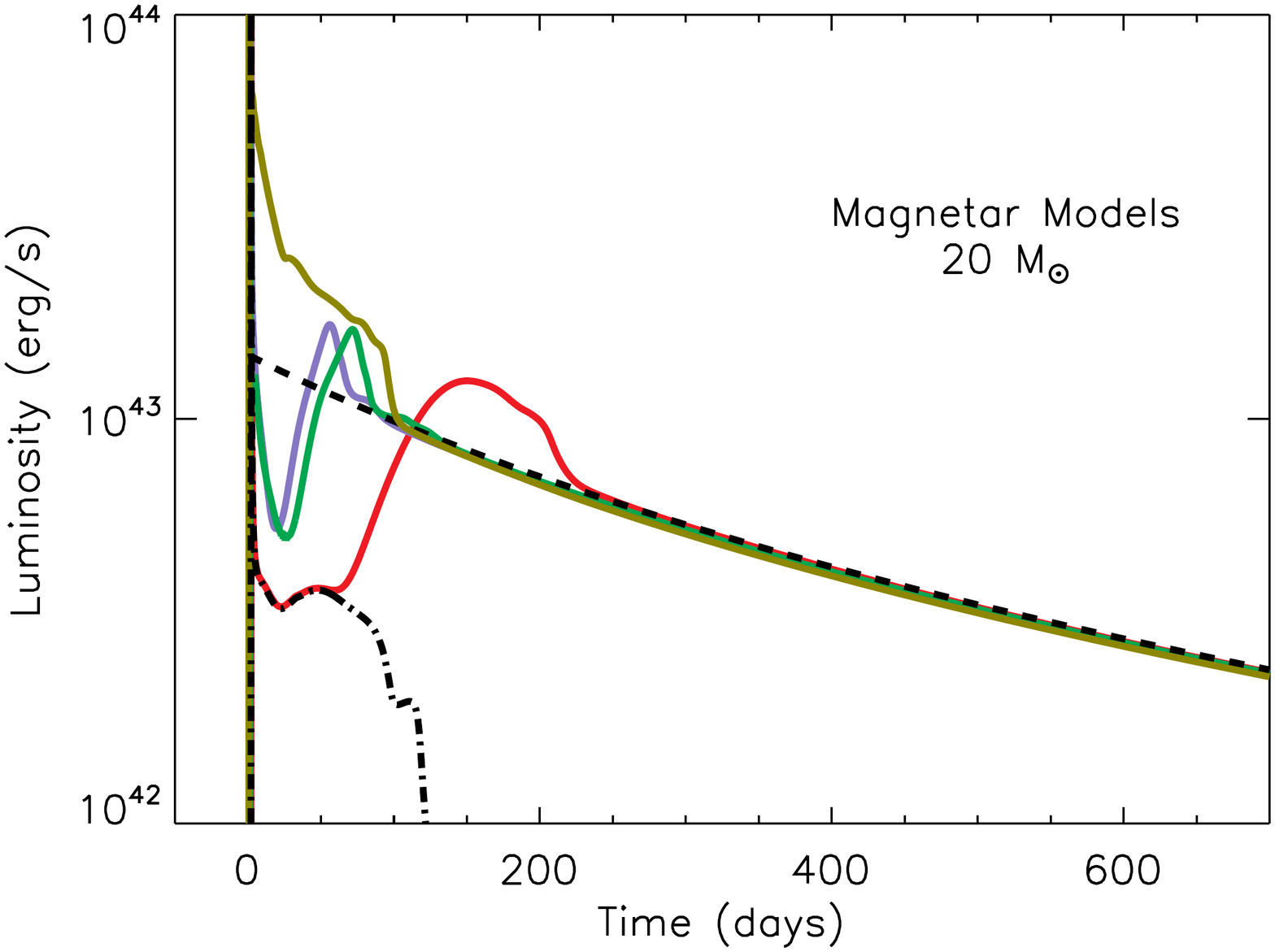}
\hfill
\includegraphics[width=0.45\textwidth]{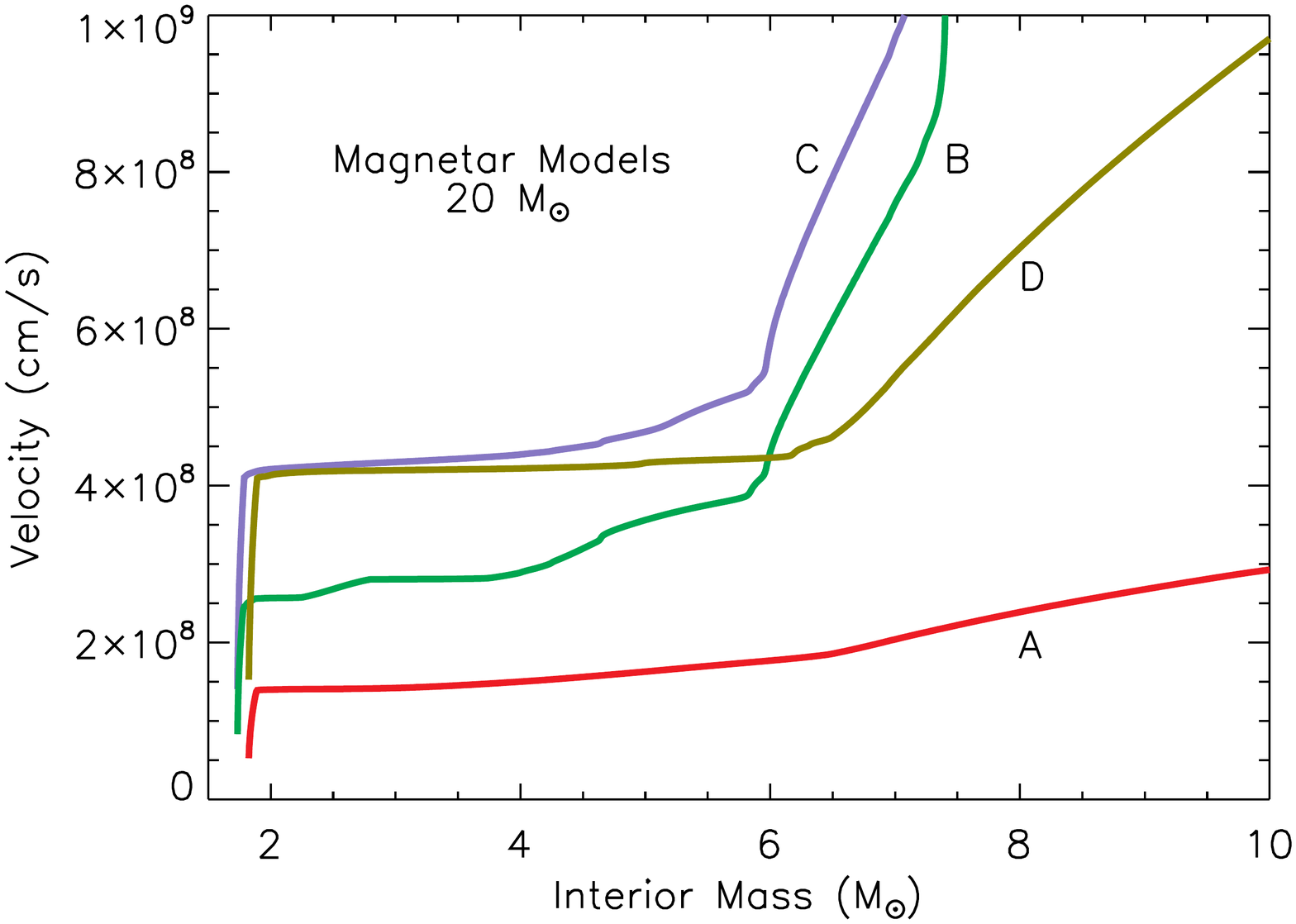}
\caption{Light curves and terminal velocities for a supernova derived
  from a 20 \Msun \ model (\Tab{models}) with several varieties of
  magnetars embedded. The brief display represented by the dot-dashed
  line in the light curve figure is a normal 20 \Msun \ supernova
  model with regular mass loss and no embedded magnetar. The dashed
  line is the dipole power radiated by a magnetar with initial
  rotational energy 0.6 $\times 10^{51}$ erg and magnetic field
  strength $4 \times 10^{13}$ G according to \eq{power}. The colored
  curves show the result of embedding this fiducial magnetar in four
  different supernovae with different mass loss histories and energy
  deposition during their first 10,000 s (see text for
  discussion). The lower panel shows the terminal velocity for the
  same four supernovae. Models A and B have a constant magnetic field
  and a normal supernova energy. Models C and D have a time varying
  field that produces a more powerful explosion early on.
  \lFig{magnetar}}
\end{center}
\end{figure}

\citet{Arc17} found a good overall fit to the light curve of iPTF14hls
using the formulae of \citet{Kas10} with B = $7 \times 10^{13}$ G and
an initial rotational period of 5 ms (E = $8 \times 10^{50}$
erg). Using \eq{power} gives a similar good fit for an initial
rotational energy of $6 \times 10^{50}$ erg and B = $4 \times 10^{13}$
G. \Fig{magnetar} and \Tab{models} show the result when a magnetar
with these properties is embedded in a supernova derived from a star
with solar metallicity and a mass of 20 \Msun \ on the main
sequence. With a normal mass loss rates, the star had a total mass
of 15.93 \Msun \ at death. The helium core mass was 6.17 \Msun \ and
the rest of the star was a low density, hydrogen-rich envelope. The
star was a a red supergiant with luminosity $5.7 \times 10^{38}$ erg
s$^{-1}$ and radius $7.42 \times 10^{13}$ cm. This star, exploded with
a piston at 1.82 \Msun \ (the base of the oxygen shell where the
entropy per baryon reaches 4.0), has been previously published in
\citet{Woo07b}. The explosion produced 0.14 \Msun \ of $^{56}$Ni and
had a final kinetic energy of $1.2 \times 10^{51}$ erg. The light
curve of this rather standard SN Type II-P model is shown as the
dot-dashed line in \Fig{magnetar}.

For Model 20A in \Fig{magnetar}, the red line with the broad peak at
150 d, is the light curve that results when power from the standard
magnetar defined above, shown as the dashed line, is embedded in this
model. Somewhat like the radioactive peak in SN 1987A, the bottled up
magnetar energy diffuses out producing a delayed peak. Once that wave
of radiation diffuses out, the bolometric light curve just reflects
the magnetar power with no delay.  Even though the bolometric light
curve is not badly represented if, say, the SN was discovered after
200 d, the low velocity shown as the red line in the second panel of
\Fig{magnetar} is much slower than the speeds observed in iPTF14hls.

Better agreement with spectral constraints is achieved if the
presupernova star has lost most of its hydrogen envelope before
exploding (Model 20B in \Fig{magnetar} and \Tab{models}). The
expansion of the helium core is then not so tamped by and the small
mass of envelope expands with a higher speed. This particular model
had a mass loss rate 3 time the standard value and ended with a total
mass of 7.41 \Msun, a helium core mass of 5.83 \Msun, and a hydrogen
envelope of 1.58 \Msun. It was exploded with a piston at 1.58 \Msun
\ and had a final kinetic energy of $1.1 \times 10^{51}$ erg. Without
an embedded magnetar, the light curve of this model (not shown) is
brief and just a bit brighter than the standard explosion, essentially
the green curve in \Fig{magnetar} before the rapid rise at $\sim25$
d. With the standard magnetar (0.6 $\times 10^{51}$ erg, $4 \times
10^{13}$ G), the light curve is similar to Model 20A, but due to the
smaller mass envelope, the bump from magnetar break out is fainter and
happens earlier. This could reduce the delay time between explosion
and discovery. The velocity in outermost part of the ejected hydrogen
envelope now exceeds the 8000 km s$^{-1}$, but only in 0.3 \Msun. It
might be difficult for this small mass to dominate the spectrum
for 600 days. The speed of the helium and heavy element core is faster
than in Model 20A, but still substantially slower than 4000 km
s$^{-1}$.

While other radii, masses of stars, explosion energies, and magnetar
properties could be explored, the properties for this one 20 \Msun
\ model are probably generic. Without greatly increasing the explosion
energy, the helium core will not move much faster. The magnetar
properties are essentially fixed by observations after break out. The
radius only affects the initial light curve and, to some extent, the
peak velocity of the ejecta. Something else may be needed to explain
the constancy of of H$_{\alpha}$ and iron speeds in the spectrum.

The missing ingredient could relate to the assumption of a constant
magnetic field strength throughout the supernova's evolution.
Consider a case where the magnetar field strength is much greater
early on and the initial rotational energy is not a perturbation on
some other undefined energy source, but actually the cause of the
explosion. During its first 100 - 1000 s, and especially its first 10
s, the magnetar will still be evolving rapidly. Damping of the initial
differential rotation and neutrinos may have already launched a
successful explosion \citep{Aki03}, but the magnetar continues to
deposit considerable energy after that adding to the kinetic energy of
the explosion. A relevant time scale for the supernova is the shock
crossing time for the helium core, about 100 s. A relevant time for
the neutron star might be the interval necessary for the crust to
form, typically estimated at minutes to hours (Sanjay Reddy, private
communication). Perhaps the neutron star forms with a powerful field
generated by convection which then decays until the crust forms and a
residual field is ``locked in''?

To explore this speculation, the explosion of the same 20 \Msun \ star
was modeled assuming a large magnetic field, $2 \times 10^{15}$ G
during the first 10$^4$ s, but only $4 \times 10^{13}$ G
thereafter. The initial rotational energy of the neutron star was
either $2 \times 10^{51}$ erg (Model 20C) or $15 \times 10^{51}$ erg
(Model 20D), but in both cases, that rotational energy has decayed to
about $6 \times 10^{50}$ erg after 10$^{4}$ s. In the high energy case
half of the initial energy was deposited in 650 s. In the low energy
case, half the energy was deposited in 5000 s. The actual time scales
are not so relevant so long as: a) most of the energy is deposited in
a few helium core expansion time scales, and b) the neutron star
retains $6 \times 10^{50}$ erg of rotational energy after a few weeks.

For the lower energy case, Model 20C, the light curve for the
progenitor with the low mass hydrogen envelope (1.58 \Msun) was very
similar to the case with constant magnetic field, Model 20B, the blue
and green light curves in \Fig{magnetar}. Doubling the final kinetic
energy does not greatly affect the light curve, but the velocity
profile is changed in an interesting way. The hydrogenic envelope
expands only slightly faster, but the helium and heavy element core
moves much faster, at a nearly constant speed close to 4000 km
s$^{-1}$. This reflects the fact that the energy deposited in the core
after the initial shock has exited is greater that from the shock
itself. As a result the entire core is compressed into a thin
shell. The inner 2 \Msun of ejecta moves with speeds between 4200 and
4400 km s$^{-1}$. By 10$^7$ s when the explosion is well into the
coasting phase, the entire helium and heavy element core (4.1 \Msun
\ of ejecta) is compressed into a shell with radius $4.2 - 5.5 \times
10^{15}$ cm moving at 4200 to 5500 km s$^{-1}$ (\Fig{magdens}). If
this shell were ionized, which it was not in the present study,
perhaps due to an overly simple treatment of radiation transport, it
would be optically thick.

% fig 11 - magnetar densities
\begin{figure}
\includegraphics[width=0.45\textwidth]{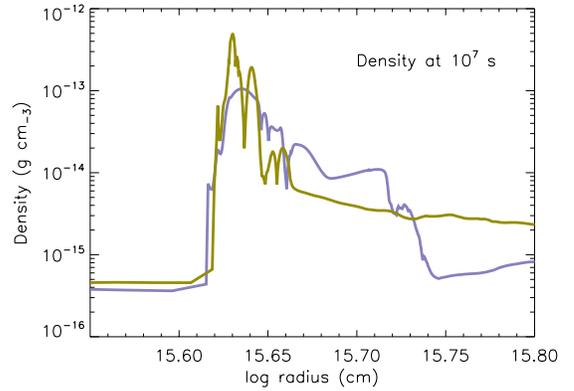}
\caption{Density at 10$^7$ s for the ejected material contained in the
  former helium and heavy element core. The blue curve is for Model
  20C and the brown curve for Model 20D. In both cases, the magnetar
  has inflated a bubble of radiation and pushed the ejected helium and
  heavy elements into a thin shell coasting with nearly constant
  velocity. Fast moving hydrogen outside has a lower density. These
  profiles scale as 1/t$^3$ with t the elapsed time greater than
  10$^6$ s.  \lFig{magdens}}
\end{figure}

Having opened the door to the possibility of a time-varying field, it
is interesting to explore the limits. \citet{Arc17} suggest that a
kinetic energy of order 10$^{52}$ erg and a mass $\sim$10 \Msun \ is
needed to explain the evolution of the Balmer lines in iPTF14hls. As
Model 20C in \Fig{magnetar} shows, even $2 \times 10^{51}$ erg is
inadequate to give the high observed speeds if progenitor explodes
with a large envelope mass.  Model 20D retains the large envelope mass
and the fiducial low energy magnetar after 10$^4$, but during the
first 10$^4$ s incorporates a magnetar with an initial rotational
energy of $15 \times 10^{51}$ erg (1.2 ms period) and magnetic field
strength $2 \times 10^{15}$ G. Now the hydrogen envelope moves faster
and now its {\sl average} speed is near 8000 km s$^{-1}$, although
with a wide spread. That part of the helium and heavy element core
that did not end up in the neutron star all moves at a nearly constant
4000 km s$^{-1}$ in a thin shell with high density contrast
(\Fig{magdens}). In the KEPLER calculation, the large radius of this
matter implies a temperature below that required to ionize helium and
the electron scattering opacity is low, but if this material had an
effective opacity near 0.1 cm$^2$ g$^{-1}$ this shell would be
marginally optically thick at 10$^{7}$ s.  Matter outside log r =
15.73 in Model 20C and log r = 15.65 for Model 20D and in
\Fig{magdens} is hydrogen rich.

In summary, given the liberty to adjust the explosion time, the
overall fit to the bolometric light curve is pretty good for all the
magnetar models, though the 1D simulations have difficulty explaining
multiple peaks in the observations. These peaks might be a consequence
of instabilities at the boundary of the ``bubble'' created inside the
expanding supernova by the magnetar radiation and wind. This boundary
is known to be Rayleigh-Taylor unstable resulting in a clumpy
filamentary structure \citep{Che16,Kas16,Blo17}. While this shell
remains optically thick, the escape of radiation could be
irregular. Alternatively the magnetar itself might experience
``superflares'' \citep{Kas17}, but, to explain a peak with integrated
luminosity $\sim10^{49}$ erg, these flares would need to be about
1,000 times more energetic than ever seen before, e.g. in the event of
March 5, 1979. Both these possibilities are speculative and would need
to be reinforced by future studies, e.g., of the light curve for a
multi-dimensional model.

A major problem though with the magnetar models considered here is
that the slowest moving hydrogen in all but the high mass, low B-field
case (Model 20A) is above 3500 km s$^{-1}$. Mixing might reduce this
value somewhat for Model 20B, but the need to have both 8000 km
s$^{-1}$ hydrogen the first 600 days, which rules out Model 20A, and
1000 km s$^{-1}$ hydrogen at late times \citep{And17}, is highly
constraining.

\section{Conclusions}
\lSect{conclude}

The most likely explanations for iPTF14hls involve CSM interaction or
magnetar birth.  For CSM interaction, the event's long duration is a
consequence the large radius, $\sim5 \times 10^{16}$ cm, of the CSM,
and relatively mild speed of the supernova ejecta. For magnetars, the
duration reflects the lengthy spin-down time for a neutron star with a
moderate field strength, $B \sim 4 \times 10^{13}$ G. 

Indeed, continuous emission lasting 600 days or more at 10$^{42}$ -
10$^{43}$ erg s$^{-1}$ (bolometric), perhaps the defining
characteristic of iPTF14hls, is easily achieved in a variety of
models, including an ordinary supernova happening in a dense CSM
medium \citep[\Sect{csm};][]{And17}; PPISN
\citep[\Sect{ppisn};][]{Arc17,And17,Woo07a,Woo17,Woo17b}; and
magnetar-based models \citep[\Sect{magnetar};][]{Arc17}. See Figures
1, 4, 5, 7, 8, and 10.  Satisfying the spectroscopic constraints is
more difficult.

Generic models for CSM interaction (\Sect{csmmodels}) require a
circumstellar mass of about 0.4 \Msun \ with an outer radius of $5
\times 10^{16}$ cm interacting with an ejected shell with 1.0 \Msun
\ and kinetic energy near $5 \times 10^{50}$ erg. The velocity of the
shock at the interface in this fiducial model would be 6500 km
s$^{-1}$ (\eq{voft}) after 600 days. At late times, the luminosity
would decline as $t^{0.3}$ (\eq{loft}) and the velocity, as
$t^{-0.1}$, but both these scaling relations depend upon 
uncertain density distributions in the ejecta and CSM and could vary
with time. Early on, one would expect narrow lines in the spectrum
characteristic of the CSM, but the pile up of matter in a dense shell
moving with the shock speed into an optically thin medium might result
in the higher velocity dominating the spectrum.  The light curve in
\Fig{csmsn} assumes a CSM density that varies smoothly as
$r^{-2}$. Irregularities in the mass loss rate or angle dependence
would be required to give structure to the light curve.

Providing a CSM of order 0.4 \Msun \ at a few $\times 10^{16}$ cm may be
difficult for binary interaction models or wave-driven mass loss. The
necessary time scale for the ejection is unnaturally short for the
former and long for the latter. An exception is a star with mass near
10 \Msun \ that ejects its hydrogen envelope a couple of years before
core collapse \citep{Woo15b}, but the explosion energies of such low
mass stars may be inadequate to produce the observed light curve
\citep{Suk16}.

The origin of distinct spectroscopic components at 4000 and 8000 km
s$^{-1}$ is not clear in the simplest spherically symmetric CSM
interaction model, though calculations of the spectra of e.g., the
model in \Fig{csmun} are needed to address this issue. One could
envision relatively minor modifications to the circumstellar mass
density \citep{And17}, mass loss history, or supernova central engine
that would give asymmetric models with the different velocities
coming from ejecta at different angles. Conceptually, despite their
uncertainties, these CSM models are the simplest explanation for
iPTF14hls.

PPISN, which are just an extreme case of CSM interaction, also have
several attractive features \citep{Arc17,And17,Woo17b}. Numerous
non-rotating models with masses 105 - 120 \Msun \ are capable of
producing continuous light curves with supernova-like brightnesses
lasting over 500, or even 1500 days (\Tab{models}; Figures 4, 5, 7,
and 8). The total energy radiated in light in these models, $2 - 3
\times 10^{50}$ erg, is the same as in iPTF14hls, Some of the models,
e.g., \Fig{t110} show variability due to multiple pulsations and
collisions with piled up material from previous pulses. Others are
capable of producing transients decades before iPTF14hls, possibly
even a transient in 1954 (\Fig{ppisnlite}; \Sect{1954}).  Each PPISN
model could, with minor modification, produce the characteristic 4000
km s$^{-1}$ seen in the iron lines of iPTF14hls. Some, with multiple
pulses, could also produce multiple spectroscopic components
(\Fig{ppisnun}). All should occur preferentially in star forming
regions with low metallicity. The physics of their explosion is simple
compared with neutrino transport or magnetar birth. They are a
phenomenon that must happen in nature provided only that stars die
with the necessary helium cores masses, 50 - 54 \Msun \ in this case.
But no one model, here at least, does it all.

In defense, PPISN are difficult to match to individual events. Their
repeated outbursts amplify small differences in the initial model. The
pair neutrino loss rate at 10$^9$ K, a relevant post-pulse
temperature, depends on temperature to the 14th power. Slight
variations in the temperature following a pulse, due e.g., to a minor
change in the amount of fuel burned in the previous pulse, have big
effects on the intervals between pulses and the light
curve. Convective mixing in the core between pulses causes some
uncertainty - when to mix and at what rate. The passage of shock waves
through the outer layers of the helium core leaves the matter that
falls back and remains bound in a state of thermal
disequilibrium. Since the temperature there is too cool for neutrino
emission, the matter remains stuck in a distended state
(\Fig{shockeddn}) that might not be accurately calculated in one
dimension. Both the interpulse period and shock hydrodynamics are
sensitive to the density distribution in this matter. For these
reasons, the interval between pulses and pulse energies were sometimes
adjusted in this paper to explore the consequences.

% fig 12 - density distriburion
\begin{figure}
\includegraphics[width=0.45\textwidth]{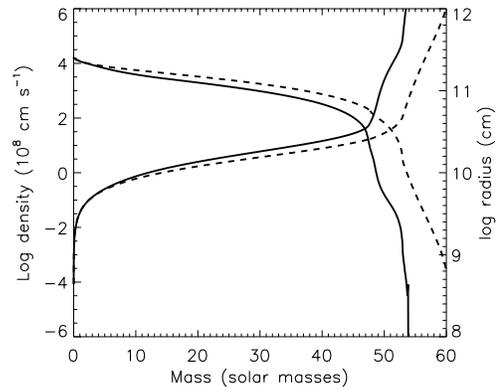}
\caption{Density distribution for Model T115 evaluated after the first
  pulse ejects all matter external to 53.9 \Msun \ (solid line) and in
  the presupernova star near carbon depletion (dashed line). The times
  have been chosen such that the central density is the same $1.6
  \times 10^4$ g cm$^{-3}$ in both cases. All matter plotted is in
  hydrostatic equilibrium and the helium core mass is 52.9 \Msun. The
  outer part of the helium core, e.g., at 50 \Msun, is disturbed by
  the first pulse and remains in an extended state of lower density and
  larger radius going into pulse 2. The entropy per baryon for the
  solid line at 50 \Msun \ (32.9) is twice that of the dashed line
  (17.3) at the same mass. The density structure of this expanded
  matter affects the interpulse period and maximum speed developed in
  pulse 2. \lFig{shockeddn}}
\end{figure}

PPISN energies are also relatively anemic. Typical total kinetic
energies for the relevant mass range are $0.7 - 1.5 \times 10^{51}$
erg (\Tab{models}), shared among two or three pulses. Even with
roughly 50\% conversion of kinetic energy to light, it is difficult
for PPISN to explain light curves totaling more than a few times
10$^{50}$ erg. The maximum kinetic energy in pulse 2 was $6.8 \times
10^{50}$ erg for Model B120 shared by 5.1 \Msun \ of ejecta.  This is
adequate for the light curve of iPTF14hls, especially if the interval
between pulses 1 and 2 can be adjusted, but too little to boost the
necessary solar mass or so of hydrogen-rich material to speeds over
8000 km s$^{-1}$. In several cases, increasing the energy of the
second pulse by less than 10$^{51}$ erg made the difference between an
acceptable model and one that lacked sufficient high velocity
hydrogen. Given the uncertainties described above, is this a
reasonable variation?

The fundamental energy limit on PPISN models, empirically, is that all
of the pulses exhaust carbon, oxygen and neon within the inner 5 \Msun
\ of the presupernova star. That is, the silicon plus iron core of a
PPISN at iron-core collapse is always near 5 \Msun \ for stars in the
relevant mass range \citep{Woo17}. Most of that is silicon. The iron
that is there is mostly made after the pulses are over. Burning a
mixture of 80\% oxygen and 20\% neon to one of 70\% silicon and 30\%
sulfur generates $4.6 \times 10^{17}$ erg g$^{-1}$, or about $5 \times
10^{51}$ erg for 5 \Msun. Most of this energy is lost to neutrinos
during the interpulse intervals, but an overall energy budget of $2
\times 10^{51}$ erg would accommodate all the artificial modifications
made in this paper.

Several varieties of PPISN models were explored, characterized by
their potential ability to explain both the 1954 transient and
iPTF14hls (\Sect{1954}); success at making only iPTF14hls
(\Sect{alone}) with a more recent unobserved supernova; and hybrid
models that invoked both a PPISN and a terminal explosion
(\Sect{hybrid}). In the first case, the interval between the beginning
of pulsations and iron-core collapse was much longer than the period
of pulsational instability. The remnant of the latest explosion would
still be a star shining with a luminosity $\sim10^{40}$ erg s$^{-1}$,
perhaps for centuries to come. This is probably too faint to detect in
iPTF14hls, and might be confused with circumstellar interaction, but
worth keeping mind for future discoveries. Other models that did not
make the 1954 transient produced a collapsed remnant, presumably a
black hole of 40 - 45 \Msun, either while the light curve was still
active or shortly thereafter. The models that made a transient in 1954
were marginally more successful at explaining both the light curve of
iPTF14hls and its high velocity hydrogen without any modification,
though none was without flaw. Models that did not attempt to make the
1954 transient had more structure and, in some cases, lasted longer,
but were less energetic, and their hydrogen-rich ejecta was slower.

The most uncertain, but potentially exciting class of PPISN models is
a hybrid model (\Sect{hybrid}) in which a PPISN is accompanied by some
sort of asymmetric terminal explosion when the iron core
collapses. This terminal explosion could possibly be magnetar
formation, which would open up a very broad parameter space, but here
the model briefly explored was black hole accretion \citep[see
  also][]{Dex13,Woo17}. The black hole would be unusual in that it
would be about 45 \Msun, and not the usual several solar masses
invoked in common supernovae or the collapsar model for gamma-ray
bursts. The matter that accretes comes from farther out in the star
and might have more angular momentum, though whether that would be
adequate to produce a jet remains to be demonstrated. This would need
to be a subset of an already rare model where the collapse to iron
core occurred within a month or so of the final pulses of the pair
instability.

Given the necessary condition of previously ejected shells within
10$^{15}$ cm being impacted by a polar outflow with equivalent
isotropic energy near 10$^{52}$ erg, a very luminous display is
generated with high characteristic speeds (\Fig{hyper}). A solid angle
of 10\% with actual explosion energy near 10$^{51}$ erg might suffice
to explain the high velocity component of iPTF14hls.

If powered by a magnetar, the light curve and kinematics of iPTF14hls
favor a star with a large energy to envelope mass ratio, i.e., Models
20B, 20C, or 20D, but not 20A (\Tab{models}). That is, the star needs
to have lost most of its hydrogen envelope or experienced a very
energetic explosion. In moderately energetic explosions, even for
stars with low mass envelopes, the mass of hydrogen moving over at
8000 km s$^{-1}$ was small (e.g., 0.3 \Msun \ in Model 20B) and might
have difficulty substantially impacting the spectrum for 600 days
\citep{Arc17}.  The presence of two characteristic speeds, 4000 km
s$^{-1}$ for the iron lines and 6000 - 8000 km s$^{-1}$ for hydrogen
is suggestive, though not proof, of two different velocity scales in
the problem. 

A novel magnetar model was explored in which the neutron star was born
rotating very fast, but with a strong magnetic field that decayed
appreciably during the first few hours of its life. The delayed energy
injection resulted in the entire ejected core of helium and heavy
elements piling up in a relatively thin shell moving with constant
speed $\sim4000$ k s$^{-1}$. Velocities in the hydrogen envelope
exceeded 8000 km s$^{-1}$ in a substantial fraction of the mass. In
any magnetar model though, the need for high velocity hydrogen during
the first 600 days and 1000 km s$^{-1}$ hydrogen at late times
\citep{And17} is problematic.

Further observations and calculations will help to clarify the actual
nature of iPTF14hls. The three models predict very different remnants
currently at the supernova site. CSM interaction in an otherwise
ordinary supernova would presumably leave an ordinary neutron star in
a shellular like remnant. PPISN predict either a Wolf-Rayet like
stellar remnant with a luminosity of 10$^{40}$ erg s$^{-1}$ or a black
hole that might be accreting and emitting hard radiation. The
most specific predictions come from the magnetar model which predicts,
three years after the explosion, a pulsar with a bolometric luminosity
of $1.3 \times 10^{42}$ erg s$^{-1}$, period, 10.5 ms, and field
strength, near $4 \times 10^{13}$ G. Five years after the explosion,
the luminosity and period would be $6.3 \times 10^{41}$ erg s$^{-1}$
and 12.3 ms. Eventually an appreciable fraction of the power should be
radiated in non-optical wavelengths. While the spectrum of a three
year old rapidly rotating magnetar is unknown, continued x-ray
monitoring at this level is recommended, as we could be witnessing the
birth of an anomalous x-ray pulsar or even a ultra-luminous x-ray
source.

In the CSM models, which include PPISN, the shock waves generating the
light should slow with time and this should be reflected in the
spectrum. PPISN models predict a characteristic speed of 1000 - 3000
km s$^{-1}$ for the envelope ejected in the first pulse (depending on
both the pulse energy and the mass of the remaining envelope). This is
intriguing given the evidence for such slowly moving material in
spectra at very late times \citep{And17}. The supernova speed would
eventually saturate near that value. In a more common supernova,
slower speeds characteristic of the pre-explosive wind might
eventually appear, but the supernova itself would retain its high speed.

In the PPISN models, the mass that was ejected consists chiefly of
hydrogen, helium, carbon, oxygen, nitrogen, neon and magnesium. Heavy
elements like silicon, calcium, and iron are confined to what existed
in the envelope of the star when it was born. No new iron or
intermediate mass elements are ejected. Even in the hybrid models with
terminal explosions, the heavy elements are presumed to collapse into
the black hole. In an ordinary supernova or magnetar model, heavy
elements would have been ejected including a solar mass or so of slowly
moving oxygen. Freshly synthesized silicon, calcium, and iron might
contribute to the spectrum.

In terms of theory, all of the models presented here would be much
more definitive with a better treatment of radiation transport.

\acknowledgements

This research has been supported by the NASA Theory Program
(NNX09AK36G). The author acknowledges illuminating exchanges with Iair
Arcavi, Jim Fuller, and Nathan Smith.

\clearpage


\vskip 0.5 in

\begin{thebibliography}{99}


\bibitem[Akiyama et al.(2003)]{Aki03} 
Akiyama, S., Wheeler, J.~C., Meier, D.~L., \& Lichtenstadt, I.\ 2003,
\apj, 584, 954

\bibitem[Andrews \& Smith(2017)]{And17} 
Andrews, J.~E., \& Smith, N.\ 2017, arXiv:1712.00514

\bibitem[Arcavi et al.(2017)]{Arc17} 
Arcavi1, U., Howell1, D. A., Kasen, D., Bildsten, L., Hosseinzadeh1,
G., et al 2017, submitted to Nature.

\bibitem[Blondin \& Chevalier(2017)]{Blo17} 
Blondin, J.~M., \& Chevalier, R.~A.\ 2017, \apj, 845, 139

\bibitem[Chen et al.(2014)]{Che14} 
Chen, K.-J., Woosley, S., Heger, A., Almgren, A., \& Whalen,
D.~J.\ 2014, \apj, 792, 28

\bibitem[Chen et al.(2016)]{Che16} 
Chen, K.-J., Woosley, S.~E., \& Sukhbold, T.\ 2016, arXiv:1604.07989
 
%\bibitem[Chen et al.(2015)]{Che15} 
%Chen,Y., Bressan, A., Girardi, L, Marigo, P., Kong, X., and Lanza, A,
%2015, \mnras, 452, 1068

\bibitem[Chevalier(1982a)]{Che82a} 
Chevalier, R.~A.\ 1982a, \apj, 258, 790 

\bibitem[Chevalier(1982b)]{Che82b} 
Chevalier, R.~A.\ 1982b, \apj, 259, 302 

\bibitem[Chevalier(1982c)]{Che82c} 
Chevalier, R.~A.\ 1982c, \apjl, 259, L85

\bibitem[Chevalier \& Fransson(1994)]{Che94} 
Chevalier, R.~A., \& Fransson, C.\ 1994, \apj, 420, 268

%\bibitem[Chevalier \& Irwin(2012)]{Che12} 
%Chevalier, R.~A., \& Irwin, C.~M.\ 2012, \apjl, 747, L17 

%\bibitem[Chugai et al.(2004)]{Chu04} 
%Chugai, N.~N., Blinnikov, S.~I., Cumming, R.~J., et al.\ 2004, \mnras,
%352, 1213

\bibitem[Dexter \& Kasen(2013)]{Dex13} 
Dexter, J., \& Kasen, D.\ 2013, \apj, 772, 30

\bibitem[Duncan \& Thompson(1992)]{Dun92} 
Duncan, R.~C., \& Thompson, C.\ 1992, \apjl, 392, L9

%\bibitem[Fern{\'a}ndez et al.(2017)]{Fer17} 
%Fern{\'a}ndez, R., Quataert, E., Kashiyama, K., \& Coughlin,
%E.~R.\ 2017, arXiv:1710.01735

\bibitem[Fuller(2017)]{Ful17} 
Fuller, J.\ 2017, \mnras, 470, 1642

\bibitem[Fuller \& Ro(2017)]{Ful17a} 
Fuller, J., \& Ro, S.\ 2017, arXiv:1710.04251 

%\bibitem[Heger, Langer, \& Woosley(2000)]{Heg00} 
%Heger, A., Langer, N., \& Woosley, S.~E.\ 2000, \apj, 528, 368

%\bibitem[Heger \& Woosley(2002)]{Heg02} 
%Heger, A., \& Woosley, S.~E.\ 2002, \apj, 567, 532

%\bibitem[Heger et al.(2005)]{Heg05} 
%Heger, A., Woosley, S.~E., \& Spruit, H.~C.\ 2005, \apj, 626, 350

%\bibitem[Heger \& Woosley(2010)]{Heg10} 
%Heger, A., \& Woosley, S.~E.\ 2011, \apj, 724, 341

\bibitem[Kasen \& Bildsten(2010)]{Kas10} 
Kasen, D., \& Bildsten, L.\ 2010, \apj, 717, 245 

\bibitem[Kasen et al.(2016)]{Kas16} 
Kasen, D., Metzger, B.~D., \& Bildsten, L.\ 2016, \apj, 821, 36

%\bibitem[Kasen et al.(2011)]{Kas11} 
%Kasen, D., Woosley, S.~E., \& Heger, A.\ 2011, \apj, 734, 102 

\bibitem[Kaspi \& Beloborodov(2017)]{Kas17}
Kaspi, V. M., \& Beloborodov, A. M. 2017, \araa, 55, 181

\bibitem[Lang(1980)]{Lan80}
Lang, K. R, 1980, {\sl Astrophysical Formulae}. (Springer Verlag),
p. 475

\bibitem[MacFadyen et al.(2001)]{Mac01} 
MacFadyen, A.~I., Woosley, S.~E., \& Heger, A.\ 2001, \apj, 550, 410

%\bibitem[Popov(1993)]{Pop93} 
%Popov, D.~V.\ 1993, \apj, 414, 712

\bibitem[Quataert \& Kasen(2012)]{Qua12} 
Quataert, E., \& Kasen, D.\ 2012, \mnras, 419, L1 

\bibitem[Quataert \& Shiode(2012)]{Qua12a} 
Quataert, E., \& Shiode, J.\ 2012, \mnras, 423, L92

\bibitem[Quataert et al.(2016)]{Qua16} 
Quataert, E., Fern{\'a}ndez, R., Kasen, D., Klion, H., \& Paxton,
B.\ 2016, \mnras, 458, 1214

\bibitem[Sukhbold et al.(2016)]{Suk16} 
Sukhbold, T., Ertl, T., Woosley, S.~E., Brown, J.~M., \& Janka,
H.-T.\ 2016, \apj, 821, 38

\bibitem[Sukhbold et al.(2018)]{Suk18} Sukhbold, T., Woosley, S., \&
  Heger, A.\ 2018, \apj, in press, arXiv:1710.03243

\bibitem[Usov(1992)]{Uso92} 
Usov, V.~V.\ 1992, \nat, 357, 472

\bibitem[Weaver, Zimmerman, \& Woosley(1978)]{Wea78} 
Weaver, T. A., Zimmerman, G. B., \& Woosley 1978, \apj, 225, 1021

%\bibitem[Woosley(1993)]{Woo93} 
%Woosley, S.~E.\ 1993, \apj, 405, 273 

\bibitem[Woosley et al.(2002)]{Woo02} 
Woosley, S.~E., Heger, A., \& Weaver, T.~A.\ 2002, Reviews of Modern
Physics, 74, 1015

%\bibitem[Woosley \& Heger(2006)]{Woo06} 
%Woosley, S.~E., \& Heger, A.\ 2006, \apj, 637, 914

\bibitem[Woosley et al.(2007)]{Woo07a} 
Woosley, S.~E., Blinnikov, S., \& Heger, A.\ 2007, \nat, 450, 390

\bibitem[Woosley \& Heger(2007)]{Woo07b} 
Woosley, S.~E., \& Heger, A.\ 2007, \physrep, 442, 269

\bibitem[Woosley(2010)]{Woo10} 
Woosley, S.~E.\ 2010, \apjl, 719, L204 

\bibitem[Woosley \& Heger(2012)]{Woo12} 
Woosley, S.~E., \& Heger, A.\ 2012, \apj, 752, 32 

%\bibitem[Woosley \& Heger(2015a)]{Woo15a} 
%Woosley, S.~E., \& Heger, A.\ 2015a, Astrophysics and Space Science
%Library, 412, 199

\bibitem[Woosley \& Heger(2015)]{Woo15b} 
Woosley, S.~E., \& Heger, A.\ 2015, \apj, 810, 34 

\bibitem[Woosley(2017a)]{Woo17}
Woosley, S. E. 2017a, \apj, 836, 244

\bibitem[Woosley(2017b)]{Woo17b} 
Woosley, S.\ 2017b, \nat, 551, 173 



\end{thebibliography}
\end{document}